\ifx\fiverm\undefined
        \newfont\fiverm{cmr5}
\fi
\input prepictex
\input pictex
\input postpictex
\catcode`\@=11
\@input{picmore.tex}

\documentstyle[11pt,epsf]{article}

\def\labelmark{}

\def\void{}
\newenvironment{formula}[1]{\def\labelname{#1}
\ifx\void\labelname\def\junk{\begin{displaymath}}
\else\def\junk{\begin{equation}\label{\labelname}}\fi\junk}%
{\ifx\void\labelname\def\junk{\end{displaymath}}
\else\def\junk{\end{equation}}\fi\junk\labelmark\def\labelname{}}
{\ifx\void\labelname\def\junk{\end{array}\end{displaymath}}
\else\def\junk{\end{array}\right.\end{equation}}
\fi\junk\labelmark\def\labelname{}\def\junk{}
}
\newenvironment{formulae}[1]{\def\labelname{#1}
\ifx\void\labelname\def\junk{\begin{displaymath}}
\else\def\junk{\begin{eqnarray}\label{\labelname}}\fi\junk}%
{\ifx\void\labelname\def\junk{\end{displaymath}}
\else\def\junk{\end{eqnarray}}\fi\junk\labelmark\def\labelname{}}

\newcommand{\beq}{\begin{formula}}
\newcommand{\eeq}{\end{formula}}
\newcommand{\beqa}{\begin{formulae}}
\newcommand{\eeqa}{\end{formulae}}
\newcommand{\eq}[1]{(\ref{#1})}
\newcommand{\nn}{\nonumber}

\newcommand{\define}{\stackrel{\mbox{def.}}{=}}
\newcommand{\dand}{\hskip-6pt &}

\newcommand{\ra}{\rightarrow}
\newcommand{\der}{\partial}
\newcommand{\eps}{\varepsilon}
\newcommand{\NP}[1]{ {\it Nucl.~Phys.} {\bf #1}}
\newcommand{\NPS}[1]{ {\it Nucl.~Phys.~(Proc.~Suppl.)} {\bf #1}} 
\newcommand{\PL}[1]{ {\it Phys.~Lett.} {\bf #1}}

\newcommand{\PR}[1]{ {\it Phys.~Rev.} {\bf #1}}
\newcommand{\PRL}[1]{ {\it Phys.~Rev.~Lett.} {\bf #1}}

\newcommand{\AP}[1]{ {\it Ann.~Phys. (N.Y.)} {\bf #1}}

\newcommand{\ACT}[1]{ {\it Acta~Phys.~Polon.} {\bf #1}}

\begin{document}
\begin{titlepage}
\setcounter{page}{1}
\renewcommand{\thefootnote}{\fnsymbol{footnote}}

\begin{flushright}
HD-THEP-00-13\\
hep-th/0003070\\
\mbox{\phantom{draft v2}} \\
\mbox{\phantom{HD-THEP-00}} \\
\end{flushright}

\vspace{5 mm}
\begin{center}
{\Large\bf Two-loop Yang-Mills theory in \\
the world-line formalism\\[1ex]
and an Euler-Heisenberg type action } 


\vspace{10 mm}
{\bf Haru-Tada Sato,
\footnote{E-mail: sato@thphys.uni-heidelberg.de }
\footnote{Present address: Theory group, KEK (Tanashi), 
Midori-machi 3-2-1, 188-8501 Tokyo, Japan}}
{\bf Michael G. Schmidt
\footnote{E-mail: m.g.schmidt@thphys.uni-heidelberg.de}}
\,\sc{and}\,
{\bf Claus Zahlten
\footnote{E-mail: zahlten@thphys.uni-heidelberg.de}}

\vspace{5mm}

\vspace{10mm}

{\it $^{}$ Institut f{\"u}r Theoretische Physik\\
Universit{\"a}t Heidelberg\\
Philosophenweg 16, D-69120 Heidelberg, Germany}
\end{center}

\vspace{18mm}
\begin{abstract}
Within the framework of the world-line formalism we write down 
in detail a two-loop Euler-Heisenberg type action for 
gluon loops in Yang-Mills theory and discuss its divergence structure.
 We exactly perform all the world-line moduli 
integrals at two loops by inserting a mass para\-meter, and 
then extract divergent coefficients to be renormalized.  
\end{abstract}

\vfill

\begin{flushleft}
\end{flushleft}

\end{titlepage}
\setcounter{footnote}{0}
\renewcommand{\thefootnote}{\arabic{footnote}}
\renewcommand{\theequation}{\thesection.\arabic{equation}}
\section{Introduction}\label{sec1}
\setcounter{section}{1}
\setcounter{equation}{0}
\indent

The Bern-Kosower method is described as a set of simple rules to 
obtain gluon scattering amplitudes at one loop , and it is known to 
improve the computational efficiency over the current Feynman diagram 
technique~\cite{BK}. Those rules are derived from a string theory in 
the limit where the inverse string tension vanishes, as a consequence 
of the idea that a string world-sheet degenerates into a desired 
particle diagram at a singular point on the boundary of moduli 
space~\cite{MT}-\cite{DIV}. The integration over moduli space 
naturally covers all necessary Feynman diagrams appearing in field 
theory, and we hence have a compact master formula for particle 
scattering amplitudes. Thus the most conspicuous point in this 
formalism is that the diagram summation is already finished in the 
formula without introducing the loop integral and the Dirac trace 
for a given scattering~\cite{FTL,five}. This idea is also applied 
to graviton scattering~\cite{gra}. 

The discovery of the Bern-Kosower rules has also stimulated 
investigations for a new mathematical structure of quantum field 
theory; how to reflect the string-like structure into field theory 
as such. The first rederivation of the Bern-Kosower rules was 
accomplished by Strassler in Ref.~\cite{St}, where the background 
field method, the proper time method and the path integral method 
for a first quantized $0+1$ dimensional field theory (world-line 
formulation) are well combined~\cite{poly}. There are many other 
fruitful examples along this stream~\cite{exam} (see Section 1 of 
Ref.~\cite{WYM} for updated references), and these examples are 
the strong incentives to study the world-line 
formalism from the theoretical point of view, especially from the 
viewpoint of its higher loop extensions~\cite{multi}  .
The present paper also discusses two-loop 
Yang-Mills theory in the world-line formalism with the aim 
to develop techniques for higher loops. 

This paper is a continuation of the previous work~\cite{WYM}, 
where the effective action of Yang-Mills theory at the two-loop order 
is derived based on the world-line formalism; also developed there 
is a certain technique, which generates multiloop generalizations of the 
one-loop trace-log (determinant) formula. However, these arguments 
are still inside the shell of formal arguments,  since we are left 
with the problem of how to deal with the multi-integrals of 
world-line moduli parameters. The moduli integrals of a higher 
genus world-sheet are too complicated to perform, while one might 
naturally expect that this situation would be improved 
in the  field theory limit. 
Although we, of course, have an option to computerize 
these complicated integrals, there are still difficulties, for example 
in three loop QED integrals~\cite{QED3}. 

It is certainly valuable to 
analyze two-loop integrals of Yang-Mills theory in the world-line 
formalism in particular if many outer particles or an Euler-Heisenberg 
type constant field are involved. It might also hint to the world-line 
moduli integrations at higher loop orders. 

In this paper, after a detailed derivation of the two-loop Euler-
Heisenberg action with gluon loops in a pseudo-abelian  gauge field 
background, we shall present some technical issues of how to 
deal with the world-line moduli integrals in the gluon effective 
action at the second order of the Taylor expansion in terms of 
external background fields. This analysis is also essential to 
examine the divergence  structure related to a wave function 
and gauge fixing parameter  renormalization. 
We only discuss the gluon loop part, since the 
ghost loop part is rather simple and can be dealt with in the 
same way as the gluon loop case. 

When we perform the integrals, we insert a mass parameter in order 
to regularize divergences. Generally speaking, massive propagators 
in the Feynman rule method are difficult to integrate in an analytic 
way. Contrastingly in our formalism, we shall go through the entire 
procedure  analytically, and all the results will be written in 
hypergeometric functions. This is certainly an intriguing point 
of this paper. 

This paper is organized as follows. 
In Section~\ref{sec2}, we write down our starting formulae for the 
gluon loop effective action at two loops. We slightly modify a few 
notations from the previous presentation~\cite{WYM} through 
the path and proper time inversions 
presented  in Appendix~\ref{ap1}. 
In Section~\ref{sec3}, a derivation of the one-loop $\beta$-function
coefficient serves as an example for how calculations in our formalism
can be simplified by specializing to the pseudo-abelian case. 
In Section~\ref{sec4}, we apply the pseudo-abelian 
technique to the calculation of the gluon effective action presented 
in Section~\ref{sec2}. Here, we only perform the world-line path 
integral parts. This Section is a completion of the parts outlined 
in the previous paper~\cite{WYM}, and the details of the computation 
are contained in Appendix~\ref{ap2}. In Section~\ref{sec5}, we 
further study how to integrate the world-line moduli integral parts, 
which are the final integrations to obtain a fully integrated form 
of Euler-Heisenberg type action. {}~For simplicity, we only consider 
the gluon kinetic term, through the Taylor expansions concerning 
the external field strength. The Taylor coefficients are the 
functions of world-line moduli parameters, and we show that these 
coefficients can be integrated; the details are in 
Appendices~\ref{ap3} and~\ref{ap4}. Appendix~\ref{ap5} is the 
Feynman diagram analysis  to be compared with our results. 

\section{Two-loop effective action}\label{sec2}
\setcounter{section}{2}
\setcounter{equation}{0}
\indent

Let us first review in brief the world-line representation
 of the two-loop effective action in Yang-Mills theory~\cite{WYM}. 
In this paper we only discuss the gluon loop part.  It is given by  
\beq{ga}
\Gamma[A]=I_1[A] + I_2[A]\ ,
\eeq
where ($I_1=\Gamma_1+\Gamma_2$, $I_2=\Gamma^{(2)}_3$ in the previous 
paper) 
\beqa{preI1}
I_1[A] &=& -\frac{1}{8} \int\limits_0^{\infty} \!\! dS
\int\limits_0^S \!\! d\tau_{\alpha} \int\limits_0^{\infty} \!\! dT_3
\oint[{\cal D}x]_S \int\limits_{w(0)=x(0)}^{w(T_3)=x(\tau_{\alpha})}
\!\!\!\!\!\!\!\!\![{\cal D}w]_{T_3} \nn\\ 
&& \times \Bigl[
\bigl(\dot{w}_{\mu}(0) - \dot{x}_{\mu}(0)\bigr)\,\dot{w}_{\rho}(T_3)\;
W^{ae}_{\mu[\sigma\rho]\nu} [x;S,\tau_\alpha,0] \nn\\
&&\hskip 20pt +\, \dot{x}_{\nu}(0) \, \dot{w}_{\rho}(T_3)\;
W^{ae}_{\mu[\sigma\rho]\mu} [x;S,\tau_\alpha,0] \Bigr]\,
W^{ea}_{\sigma\nu}[w;T_3,0] \ ,
\eeqa
and
\beqa{I2} 
I_2[A] &=& \frac{1}{4} \int\limits_0^{\infty}\!dT_1\,dT_2
  \oint [{\cal D}x_1]_{T_1} \oint [{\cal D}x_2]_{T_2}\;
  \delta^4\!\left(x^\mu_1(0)-x^\mu_2(0)\right) \nn\\
&& \hspace{7mm} \times\, \mbox{Tr}_C
   \Bigl[\lambda^a W_{\mu\nu}[x_1;T_1,0] \Bigr] 
   \,\mbox{Tr}_C 
   \Bigl[ \lambda^a W_{\nu\mu}[x_2;T_2,0] \Bigr]\ ,
\eeqa 
with the following compact notations:
\beq{dx}
\int[{\cal D}x]_T\,F[x] = \int{\cal D}x\,e^{-{1\over4}\int\limits_0^T
{\dot x}^2\, d\tau} F[x] \qquad\mbox{for any functional}\ F[x]\ ,
\eeq
\beq{W}
W^{ea}_{\sigma\nu}[w;T_3,0]=
  \mbox{P}\exp\left\{  \int\limits_0^{T_3} 
d\tau M_{\tau}[w]\right\}^{ea}_{\sigma\nu}   \ ,
\eeq
\beq{defW}
W^{ae}_{\mu\sigma\rho\nu}[x;S,\tau_\alpha,0]=
\mbox{Tr}_C\biggl[  \lambda^a \,\mbox{P}\exp\left\{
\int\limits_{\tau_\alpha}^S d\tau M_{\tau}[x]\right\}_{\mu\sigma}
\!\!\!\!\!  \lambda^e \,\mbox{P}\exp\left\{ 
\int\limits_0^{\tau_\alpha} d\tau M_{\tau}[x]\right\}_{\rho\nu}
\biggr] \ ,  \eeq
and 
\beq{defM}
\left(M_{\tau}[x]\right)^{ab}_{\mu\nu} =
2i\left[F^c_{\mu\nu}(x(\tau^{\prime})) -\frac{1}{2}\delta_{\mu\nu}
A^c_{\rho}(x(\tau^{\prime}))\cdot\partial_{\tau^{\prime}} 
x^{\rho}(\tau^{\prime})
\right]_{\tau^{\prime}=\tau} (\lambda^c)^{ab}\ .
\eeq
Here we have slightly changed the notations used in the 
previous paper~\cite{WYM}; (i) the previous definition of $M$ 
is associated  with $D=\der-iA$,  while the 
present $M_\tau$ is associated with $D=\der+iA$, 
(ii) the path ordering directions are modified to be the standard 
one, and some related formulae are listed in Appendix~\ref{ap1}. 
As in the previous paper, we always use Euclidean space-time conventions.

\vspace{8mm}
\begin{minipage}[htb]{15cm} 
\begin{center}
\input{sym.pictex}
\end{center}
{\bf Figure 1:} (a) The loop type parametrization. 
(b) The symmetric parametrization. 
\end{minipage}
\vspace{8mm}

It is convenient to have another representation for $I_1[A]$, 
based on the symmetric parametrization  in Figure 1(b), 
which treats the individual gluon lines in a more equal
way, thus allowing a greater class of transformations by inverting 
and relabelling the gluon paths and leading to significant 
simplifications in the concrete calculations. 
The expression \eq{preI1} is based on the loop type 
parametrization  in Figure 1(a). The transformation rules 
between (a) and (b) 
are known~\cite{multi,HTS}, and we thus have the following symmetric 
representation for $I_1[A]$:
\beqa{I1} 
\lefteqn{ I_1[A]  =  - \frac{1}{8} \int\limits_0^{\infty} 
\!\! dT_1 dT_2 dT_3 \int\!\!d^D\!y_1 \int\!\!d^D\!y_2\;
\Biggl[ \prod_{k=1}^3 \int\limits_{x_k(0)=y_1}^{x_k(T_k)=y_2}
\!\!\!\!\!\!\!\!\![{\cal D}x_k]_{T_k} \Biggr]} \nn\\
&& \times \Bigl[\; 
\bigl(\dot{x}_{3\mu}(0) - \dot{x}_{1\mu}(0)\bigr)\,
\dot{x}_{3\rho}(T_3)\; 
{\tilde W}^{ae}_{\mu[\sigma\rho]\nu} [x_2^{-1},x_1;T_2,0,T_1,0] \nn\\
&&\ \ \ +\, \dot{x}_{1\nu}(0)\,\dot{x}_{3\rho}(T_3)\;
{\tilde W}^{ae}_{\mu[\sigma\rho]\mu} [x_2^{-1},x_1;T_2,0,T_1,0] 
\;\Bigr]\,W^{ea}_{\sigma\nu}[x_3;T_3,0]\ ,
\eeqa
where 
\beq{TTS}
T_1=\tau_\alpha\ ,\qquad T_2=S-\tau_\alpha \ ,
\eeq
and
\beq{Wae}
{\tilde W}^{ae}_{\mu\sigma\rho\nu} [x_2^{-1},x_1;T_2,0,T_1,0] =
\mbox{Tr}_C\biggl[
  \lambda^a \,\mbox{P}\exp\left\{ \int\limits_0^{T_2}
     \!\!d\tau M_{\tau}[x_2^{-1}]\right\}_{\mu\sigma}
  \!\!\!\!\!\lambda^e \,\mbox{P}\exp\left\{                   
  \int\limits_0^{T_1} \!\!d\tau M_{\tau}[x_1]\right\}_{\rho\nu}
  \biggr] \ . 
\eeq
Note that the transition from loop type to symmetric parametrization
requires both splitting the loop path into two parts and
inverting one of them (which we denote by $x_2^{-1}$)
to achieve all three paths to start at $y_1$ and to end at $y_2$.
In contrast to the naive expectation, 
this suggests that in a general background, one may not
just write down the product of three propagators 
starting and ending at identical points to
represent a loop with inserted propagator.
See Appendix~\ref{ap1} for a detailed definition of the 
notation $x_2^{-1}$ and some comments on path inversion 

After finishing the next section, we shall discuss the world-line 
path integrals for $x_k$ in 
Section~\ref{sec4}, and the integrals of the world-line moduli 
parts (proper times) $T_k$ in Section~\ref{sec5}. 

\section{The pseudo-abelian case }\label{sec3}
\setcounter{section}{3}
\setcounter{equation}{0}
\indent

For the rest of this paper, we confine ourselves to the 
pseudo-abelian su(2) with constant field strength. Thus we assume
\begin{equation} \label{decomp}
  A^a_{\mu}(x)  =  {\cal A}_{\mu}(x)\, n^a  
  \qquad\mbox{with}\quad n^a n^a = 1 \quad\mbox{and constant}\; n\ ,
\end{equation}
i.e.~the color dependence of the non-abelian gauge fields is supposed
to be factored out in form of a constant unit vector in color space.
Within these settings calculations are simplified considerably, though
non-abelian results can still be reproduced as shall be seen below.
As an example and to introduce some notations, we here show a brief 
sketch of how our formalism works within the calculation of 
one-loop $\beta$-function coefficients.

The assumption (\ref{decomp}) leads to similar decompositions for the
field strength 
\begin{equation}
  F^a_{\mu\nu}(x) = {\cal F}_{\mu\nu}(x)\,n^a 
  \quad\mbox{with}\quad
  {\cal F}_{\mu\nu}(x)=
  \partial_{\mu}{\cal A}_{\nu}(x)-\partial_{\nu}{\cal A}_{\mu}(x)
\end{equation}
and the matrix $M_\tau[x]$
\begin{equation}
  M_{\tau}[x] = {\cal M}_{\tau}[x]\otimes(n^c \lambda^c)
  \equiv {\cal M}_{\tau}[x]\otimes{\cal T}_- \ ,
\end{equation}
where ${\cal M}_{\tau}[x]$ is the Lorentz matrix defined by
\begin{equation} \label{calMtauA}
  {\cal M}_{\tau}[x]
  = 2i \Bigl[
  {\cal F}(x)-\frac{1}{2}
  \,{\cal A}_{\rho}(x)\,\dot{x}_{\rho}\,\bf{1}_L
  \Bigr]  
\end{equation}
and ${\bf 1}_L$ denotes the unit Lorentz matrix (in the Euclidean space).

So far we have not used our additional assumption of a constant field 
strength, nor have we fixed
the gauge for the external gauge fields $A^a_\mu$. If we take into account
the constancy of the field strength, we may choose
\begin{equation}
  {\cal A}_{\mu}(x) = \frac{1}{2}x_{\nu}{\cal F}_{\nu\mu}\ ,
\end{equation}
henceforth expecting ${\cal M}_{\tau}[x]$ to be of the form
\beq{calMtau}
{\cal M_{\tau}}[x] = 2i \Bigl[{\cal F}-\frac{1}{4}\,
x_\sigma {\cal F}_{\sigma\rho}\,\dot{x}_{\rho}\,{\bf 1}_L \Bigr] , 
\eeq
rather than (\ref{calMtauA}). 
In addition, it is convenient to define 
the integrated matrices (omitting the index $\tau$),
and we can simply write
\beq{Mpseudo}
\int\limits_0^T M_{\tau}[x] d\tau = {\cal M}[x]\otimes{\cal T}_- 
\qquad\mbox{with}\quad
{\cal M}[x] = \int\limits_0^T {\cal M}_{\tau}[x] d\tau
\ .
\eeq

It is the benefit of confining ourselves to the pseudo-abelian case, that
the $M_\tau[x]$ matrices for different values of the parameter $\tau$ 
become commuting quantities
\beq{Mcomm}
[\,M_{\tau}[x],\,M_{\tau^{\prime}}[x]\,] = 0 \ .
\eeq
Thus we are allowed to drop path ordering from all of our expressions. 
This leads  to the following decomposition
\beq{**}
\mbox{P}\exp\left\{ {\cal M}\otimes{\cal T}_- \right\}
= {\bf 1}_L \otimes {\cal I} \;+\;
\sinh\left\{ {\cal M}\right\} \otimes{\cal T}_- 
\;+\; \cosh\left\{{\cal M}\right\} \otimes{\cal T}_+\ , 
\eeq
in terms of the $su(2)$ matrices
\beq{su2}
{\cal T}_- = n^c \lambda^c\ ,\quad
{\cal T}_+ = ({\cal T}_-)^2\ ,\quad
{\cal I} = {\bf 1}_C - {\cal T}_+\ , \qquad
\mbox{where} \quad {\bf 1}_C=diag(1,1,1)\ .
\eeq
Now, using the properties
\beq{trsu2}
\mbox{Tr}_C\,{\cal T}_-=0,\qquad
C_A\equiv \mbox{Tr}_C \,{\cal T}_+=2,\qquad
\mbox{Tr}_C\,{\cal I}=1 
\eeq
the one-loop effective action for the gluon loop is calculated 
as follows~\cite{RSS}:
\beqa{Gas}
\Gamma^{1-loop}_G[A] &=& -{1\over2}\int\limits_0^\infty{dT\over T}
e^{-m^2T}\oint[{\cal D}x]_T\, 
\Bigl(\mbox{P}\exp\int\limits_0^T M_\tau[x]\,d\tau \Bigr)^{aa}_{\mu\mu}\nn\\
&=&-{1\over2}\int\limits_0^\infty{dT\over T}e^{-m^2T}
\oint[{\cal D}x]_T\,
\left( 
D + C_A \mbox{Tr}_L(\cosh{\cal M}) \right)\nn\\
&=&-{1\over4}\int\limits_0^\infty{dT\over T}e^{-m^2T}
\oint[{\cal D}x]_T\,
\left( 
D + C_A \mbox{Tr}_L(e^{\cal M}) \right)
\quad + \quad  ({\cal F}\rightarrow -{\cal F})\ ,
\eeqa
where we have introduced the gluon mass term $e^{-m^2T}$ for regularization.
The second contribution with ${\cal F}$ replaced by ${-\cal F}$ counts
for a factor of two, thus with the one-loop path integral normalization
\beq{oneloopint}
\oint{\cal D}x\, \exp\biggl\{-{1\over4}\int\limits_0^T \!\!d\tau
\left[
{\dot x}^2 +2i x{\cal F}{\dot x} 
\right]
\biggr\}
=(4\pi T)^{-D/2}\mbox{det}_L^{-1/2}\Bigl(\,
{\sin{\cal F}T\over{\cal F}T}\,\Bigr)
\int\!\! d^D\!x_0 
\eeq
we are led to
\begin{equation}
  \label{eq:G-1loop}
  \Gamma^{1-loop}_G[A]=
  \frac{-1}{2(4\pi)^{D/2}}\int\limits_0^\infty\!\!dT T^{-1-D/2} e^{-m^2 T}
  \left[
    D + C_A \mbox{Tr}_L(e^{2i{\cal F}T})
    \,\mbox{det}_L^{-1/2}\Bigl(\,
    {\sin{\cal F}T\over{\cal F}T}\,\Bigr)
  \right]
  \int\!\! d^D\!x_0\ .
\end{equation}
For now we are interested in the two-point function only, i.e.~in the second
functional derivative of the effective action with respect to $A_{\mu}^a$. 
To this end, we only need 
the second order term of an expansion of (\ref{eq:G-1loop}) in terms 
of ${\cal F}$. Omitting constant and higher order terms we find
\beq{gf2}
\Gamma^{1-loop}_G[A] = \ \cdots -\ \frac{C_A}{2(4\pi)^{D/2}}
\Bigl({D\over12}-2\Bigr)
\int\limits_0^\infty\!\! dT\,T^{1-D/2}\,
e^{-m^2T} \int\!\!d^D\!x_0\,\mbox{Tr}_L{\cal F}^2 \ +\cdots \ .
\eeq
Setting $D=4-2\eps$ and performing the $T$ integration leads to the 
following pole structure in $\eps$:
\beq{...}
\Gamma_G^{1-loop}[A]=\ \cdots -\ {g_0^2 C_A\over(4\pi)^2}
\Biggl({-10\over3\eps} \Biggr)
\int\!\!d^D\!x_0\,
\Biggl(-\frac{1}{4}{\cal F}_{\mu\nu}{\cal F}_{\mu\nu} \Biggr)
+{\cal O}(\eps^0)\ +\cdots\ ,
\eeq 
where we have revived the gauge coupling $g$ and where  
$g_0$ is the dimensionless coupling constant defined by 
$g=g_0\mu^\eps$.
Finally we calculate the functional derivative and transform into momentum 
space: using 
\begin{equation} \label{FT}
  \int\!\!d^D\!x_0\,
  \Biggl(-\frac{1}{4}{\cal F}_{\rho\sigma}{\cal F}_{\rho\sigma} \Biggr)
  \;\longrightarrow\;
  - \delta^{ab}(\delta_{\mu\nu}k^2 - k_{\mu}k_{\nu})
\end{equation}
we read off
\beq{PiG}
{\Pi_G}^{ab}_{\mu\nu}= -\frac{g_0^2 C_A \delta^{ab}}{(4\pi)^2}
\Biggl({10\over3\eps} \Biggr)
(\, \delta_{\mu\nu}k^2 - k_\mu k_\nu \,) +{\cal O}(\eps^0)\ .
\eeq

Similarly, the one-loop contribution from a ghost loop can 
be calculated:
As can be deduced from Ref.~\cite{WYM}, the ghost one-loop action is 
given by changing the overall normalization in \eq{Gas} from 
$1/2$ to $-1$, and only employing the Lorentz scalar term 
in \eq{defM}; i.e., define $-iA^c_\mu{\dot x}_\mu\lambda^c 
\equiv {\tilde M}_\tau$ instead of using $M_\tau$. 
The corresponding pseudo-abelian quantity $\tilde{\cal M}$ is 
defined analogically to the gluon loop case 
(q.v. Eqs.~\eq{calMtau} and \eq{Mpseudo}). 
Thus we find the one-loop ghost action
\beqa{.1}
\Gamma^{1-loop}_{FP}[A] &=& 
\int\limits_0^\infty{dT\over T}
e^{-m^2T}\oint[{\cal D}x]_T\, 
\Bigl(\mbox{P}\exp\int\limits_0^T \tilde{M}_\tau[x]\,d\tau \Bigr)^{aa}\nn\\
&=&
\int\limits_0^\infty{dT\over T}e^{-m^2T}
\oint[{\cal D}x]_T\,
\left( 1 + C_A \cosh\tilde{\cal M} \right)\nn\\
&=&
{1\over2}\int\limits_0^\infty{dT\over T}e^{-m^2T}
\oint[{\cal D}x]_T\,
\left( 1 + C_A e^{\tilde{\cal M}} \right)
\quad + \quad  ({\cal F}\rightarrow -{\cal F})\ .
\eeqa
Again taking into account the $-{\cal F}$ term by a factor of two and 
using \eq{oneloopint}, we arrive at
\begin{equation}
  \label{eq:FP-1loop}
  \Gamma^{1-loop}_{FP}[A]=
  \frac{1}{(4\pi)^{D/2}}\int\limits_0^\infty\!\!dT\,T^{-1-D/2} e^{-m^2 T}
  \left[
    1 + C_A \mbox{det}_L^{-1/2}\Bigl(\,
    {\sin{\cal F}T\over{\cal F}T}\,\Bigr)
  \right]
  \int\!\! d^D\!x_0\ .
\end{equation}
Expanding this expression in the same way as done in \eq{gf2}, we derive
\beqa{.2}
\Gamma^{1-loop}_{FP}[A] &=& 
\ \cdots +\frac{C_A}{12(4\pi)^{D/2}}
\int\limits_0^\infty\!\! dT\,T^{1-D/2}\,
e^{-m^2T} \int\!\!d^D\!x_0\,\mbox{Tr}_L{\cal F}^2 \ +\cdots\\
  &=& 
\ \cdots -{g_0^2 C_A\over(4\pi)^2}
\Biggl({-1\over3\eps} \Biggr)
\int\!\!d^D\!x_0\,
\Biggl(-\frac{1}{4}{\cal F}_{\mu\nu}{\cal F}_{\mu\nu} \Biggr)
+{\cal O}(\eps^0)\ +\cdots\ ,
\eeqa
and hence
\beq{PiFP}
{\Pi_{FP}}^{ab}_{\mu\nu}= -{g_0^2 C_A \delta^{ab}\over(4\pi)^2}
\Biggl({1\over3\eps} \Biggr)
(\,\delta_{\mu\nu}k^2 - k_\mu k_\nu \,) +{\cal O}(\eps^0)\ .
\eeq
Gathering Eqs.~\eq{PiG} and \eq{PiFP}, the correct (one-loop) 
$\beta$-function coefficient $11/3$ is reproduced.

\section{Two-loop Euler-Heisenberg formulas}\label{sec4}
\setcounter{section}{4}
\setcounter{equation}{0}
\indent

Now, let us consider the extension of the above calculations to the 
two-loop case. We deal with the symmetric representations \eq{I2} 
and \eq{I1}. In the two-loop case , as understood from \eq{defM}, we 
have to keep in mind that the 
sign  of the $x{\cal F}{\dot x}$ 
term changes due to the $\tau$ derivative, 
if we invert the path. 
For example in \eq{Wae}, 
one should notice (see also Appendix~\ref{ap1}) that 
\begin{eqnarray} \label{MT_detail}
\Bigl(M_{\tau}[x_2^{-1}]\Bigr)^{ab}_{\mu\nu} & = &
2i\left[F^c_{\mu\nu}(x_2^{-1}(\tau^{\prime})) -\frac{1}{2}\delta_{\mu\nu}
A^c_{\rho}(x_2^{-1}(\tau^{\prime}))\cdot\partial_{\tau^{\prime}} 
x_{2\rho}^{-1}(\tau^{\prime})
\right]_{\tau^{\prime}=\tau} (\lambda^c)^{ab}\nonumber\\
& = &
2i\left[F^c_{\mu\nu}(x_2(\tau^{\prime})) +\frac{1}{2}\delta_{\mu\nu}
A^c_{\rho}(x_2(\tau^{\prime}))\cdot\partial_{\tau^{\prime}} 
x_{2\rho}(\tau^{\prime})
\right]_{\tau^{\prime}=T_2-\tau} (\lambda^c)^{ab}\nonumber\\
& = &
\Bigl(M_{T_2-\tau}[x_2]\Bigr)^{ba}_{\nu\mu}\ .
\end{eqnarray}
Reflecting this fact, it is rather convenient to introduce 
the signature index $\kappa$ $(=\pm1,0)$ on the Lorentz matrix
${\cal M}$ :
\beq{Mk}
{\cal M}^{(\kappa)}_k \define \int\limits_0^{T_k}\!\!d\tau\,
2i \Bigl[   |\kappa|{\cal F}-\frac{1}{4}\;\kappa \;
x_{k\sigma}{\cal F}_{\sigma\rho}\, \dot{x}_{k\rho}\,{\bf 1}_L
\Bigr]\ ,
\eeq
where the $k$ stands for the line labels 1,2 and 3. 
With this matrix notation \eq{Mk}, the general form for the color 
matrix part of the action \eq{I1} is written in the form 
\beqa{trW} 
\lefteqn{
{\tilde W}^{ae}_{\gamma\delta\alpha\beta}[x_2^{-1},x_1;T_2,0,T_1,0]\,
W^{ea}_{\sigma\nu}[x_3;T_3,0]}  \nn \\
&=& \mbox{Tr}_C\biggl[ \lambda^a \, \exp\left\{ 
{\cal M}^{(-)}_2 \otimes{\cal T}_- \right\}_{\gamma\delta} 
\lambda^e \, \exp\left\{  {\cal M}^{(+)}_1 \otimes
{\cal T}_- \right\}_{\alpha\beta} \biggr] 
\,\exp\left\{ {\cal M}^{(+)}_3 \otimes{\cal T}_ 
- \right\}^{ea}_{\sigma\nu} \ .
\eeqa
After using the expansion (cf.~Eq.~(\ref{**}))
\beq{expM}
\exp\left\{ {\cal M}^{(\kappa)}_k \otimes{\cal T}_- \right\}
= {\bf 1}_L \otimes {\cal I} \;+\;
\sinh\left\{ {\cal M}^{(\kappa)}_k\right\} \otimes{\cal T}_- 
\;+\; \cosh\left\{
{\cal M}^{(\kappa)}_k\right\} \otimes{\cal T}_+\ , 
\eeq
we perform the color traces  applying the following formulae: 
\beq{.5}
\begin{array}{lcl}
  \mbox{Tr}_C \left(
  \lambda^a {\cal T}_{\pm}\,\lambda^e {\cal T}_{\pm}\right) 
  {\cal I}^{ea} & = & 2 \\
  \mbox{Tr}_C \left(
  \lambda^a {\cal T}_{\pm}\,\lambda^e {\cal I}\right) 
  {\cal T}^{ea}_{\pm} & = & 2 \\
  \mbox{Tr}_C \left(
  \lambda^a {\cal I}\,\lambda^e {\cal T}_\pm\right) 
  {\cal T}^{ea}_\pm & = & \pm 2 \ ,
\end{array}
\eeq
where the 3rd formula follows from the 2nd one with the properties
\beq{.6}
({\cal T}_\pm)^T=\pm{\cal T}_\pm, \qquad
{\cal I}^T={\cal I} \ . 
\eeq
{}~For any other combinations of $A$, $B$ and $C$ chosen out of  
$\{{\cal I},\;{\cal T}_-,\;{\cal T}_+\}$, the 
following formula applies 
\beq{.7}
\mbox{Tr}_C \left(\lambda^a A\,\lambda^e B\right) C^{ea} = 0\ .
\eeq
Thus the quantity~\eq{trW} is calculated 
as follows:
\beqa{(*)}
\lefteqn{
{\tilde W}^{ae}_{\gamma\delta\alpha\beta}[x_2^{-1},x_1;T_2,0,T_1,0]\,
W^{ea}_{\sigma\nu}[x_3;T_3,0] }  \nn \\
&=& \hspace{9.5pt}   2\,\delta_{\alpha\beta}\biggl[
\cosh\left\{{\cal M}^{(-)}_2\right\}_{\gamma\delta}
\cosh\left\{{\cal M}^{(+)}_3\right\}_{\sigma\nu} +\;
\sinh\left\{{\cal M}^{(-)}_2\right\}_{\gamma\delta}
\sinh\left\{{\cal M}^{(+)}_3\right\}_{\sigma\nu}\,\biggr]\nn\\
&& +\,
  2\,\delta_{\gamma\delta}\,\biggl[
  \cosh\left\{{\cal M}^{(+)}_1\right\}_{\alpha\beta}
  \cosh\left\{{\cal M}^{(+)}_3\right\}_{\sigma\nu} \!-\;
  \sinh\left\{{\cal M}^{(+)}_1\right\}_{\alpha\beta}
  \sinh\left\{{\cal M}^{(+)}_3\right\}_{\sigma\nu}
  \biggr] \nn \\
&& +\, 
  2\,\delta_{\sigma\nu} \hspace{1.0pt} \biggl[
  \cosh\left\{{\cal M}^{(+)}_1\right\}_{\alpha\beta}
  \cosh\left\{{\cal M}^{(-)}_2\right\}_{\gamma\delta} \! +\;
  \sinh\left\{{\cal M}^{(+)}_1\right\}_{\alpha\beta}
  \sinh\left\{{\cal M}^{(-)}_2\right\}_{\gamma\delta}\,\biggr]\\
&=& \hspace{11pt}
  \exp\left\{ {\cal M}^{(0)}_1\right\}_{\alpha\beta}
  \exp\left\{ {\cal M}^{(-)}_2\right\}_{\gamma\delta}
  \exp\left\{ {\cal M}^{(+)}_3\right\}_{\sigma\nu}
  \nn \\
&& +\,
  \exp\left\{ {\cal M}^{(+)}_1\right\}_{\alpha\beta}
  \exp\left\{ {\cal M}^{(0)}_2\right\}_{\gamma\delta}
  \exp\left\{ -{\cal M}^{(+)}_3\right\}_{\sigma\nu}
  \nn \\
&& +\,
  \exp\left\{ {\cal M}^{(+)}_1\right\}_{\alpha\beta}
  \exp\left\{ {\cal M}^{(-)}_2\right\}_{\gamma\delta}
  \exp\left\{ {\cal M}^{(0)}_3\right\}_{\sigma\nu}
  \quad + \quad  ({\cal F}\rightarrow -{\cal F})\ .
\eeqa
Now, as sketched in Ref.~\cite{WYM}, performing the trivial integral 
(the 1st term) in Eq.\eq{Mk} 
\beqa{.8}
\lefteqn{
\exp\Bigl\{\int\limits_0^{T_k}\!\!d\tau
\bigl(-\frac{1}{4}\,\dot{x}_k^2\bigr)\Bigr\}
\exp\left\{\pm{\cal M}^{(\kappa)}_k\right\}_{\eta\xi}}\nn\\
& = &
\exp\left\{\pm 2i\,|\kappa|\,T_k\,{\cal F}\right\}_{\eta\xi}
\exp\Bigl\{-\frac{1}{4}\int\limits_0^{T_k}\!\!d\tau
\bigl[\dot{x}_k^2 + 2i\,(\pm\kappa)\,  x_{k\sigma}
{\cal F}_{\sigma\rho}\,\dot{x}_{k\rho}\bigr]\Bigr\}\ ,
\eeqa
and introducing the quantities 
\beq{action3}
  S^{(\kappa_1,\,\kappa_2,\,\kappa_3)} =
  -\frac{1}{4}\sum_{k=1}^3 \int\limits_0^{T_k}\!\!d\tau
  \Bigl[\,\dot{x}_k^2 + 2i\,\kappa_k\,  x_{k\sigma}
{\cal F}_{\sigma\rho}\,\dot{x}_{k\rho} \,\Bigr] \ ,
\qquad (\kappa_a=\pm1,0)
\eeq
we have the following formula:
\beqa{Wform}
& & \!\!\!\!\!\!\!\!\!\!\!\!\!\!\!
\exp\Bigl\{-\frac{1}{4}\sum_{k=1}^3 \int\limits_0^{T_k}\,d\tau
  \,\dot{x}_k^2\Bigr\} \,
{\tilde W}^{ae}_{\gamma\delta\alpha\beta}[x_2^{-1},x_1;T_2,0,T_1,0]\,
W^{ea}_{\sigma\nu}[x_3;T_3,0]
\nn\\
& = & \hspace{8.0pt}  \delta_{\alpha\beta}\;
  \exp\Bigl\{2i\,T_2\,{\cal F}\Bigr\}_{\gamma\delta}
  \exp\Bigl\{2i\,T_3\,{\cal F}\Bigr\}_{\sigma\nu}
  \hspace{8.0pt} e^{S^{(0,-,+)}}   \nn\\
&  & +\,  \delta_{\gamma\delta}\;
  \exp\Bigl\{2i\,T_1\,{\cal F}\Bigr\}_{\alpha\beta}
  \exp\Bigl\{2i\,T_3\,{\cal F}\Bigr\}_{\nu\sigma}
  \hspace{8.0pt} e^{S^{(+,0,-)}}  \nn\\
&  & +\,  \delta_{\sigma\nu}\;
  \exp\Bigl\{2i\,T_1\,{\cal F}\Bigr\}_{\alpha\beta}
  \exp\Bigl\{2i\,T_2\,{\cal F}\Bigr\}_{\gamma\delta}
  \hspace{8.0pt} e^{S^{(+,-,0)}}
  \quad + \quad ({\cal F}\rightarrow -{\cal F})\ .
\eeqa
It is worth noticing here that the interaction terms 
$x_k{\cal F}{\dot x}_k$ defined on the three different  lines possess 
different $\kappa$ values; this fact 
is  related to the 
$su(2)$ structure $\varepsilon^{abc}$. In the following, we 
calculate $I_1$ and $I_2$ separately, since these two quantities 
involve different world-line topology. 

\subsection{The $I_1[A]$ part}\label{sec41}
\indent

Applying the formula \eq{Wform} to Eq.~\eq{I1}, we obtain the 
following expressions (For the convenience of presentation, we 
split $I_1[{\cal A}]$  into two quantities depending on whether 
$x_1x_3$ or $x_3x_3$ correlations.): 
\beq{.9}
I_1[{\cal A}]=\Gamma_1[{\cal A}]+\Gamma_2[{\cal A}]\ ,
\eeq
where
\beqa{gamma1}
\lefteqn{
\Gamma_1[{\cal A}] = -\frac{1}{8} 
\int\limits_0^{\infty} 
\!\! dT_1 dT_2 dT_3 \int\!\!d^D\!y_1  \int\!\!d^D\!y_2 \;
\Biggl[  \prod_{k=1}^3 \int\limits_{x_k(0)=y_1}^{x_k(T_k)=y_2}
\!\!\!\!\!\!\!\!\!{\cal D}x_k  \Biggr] 
\dot{x}_3^\mu(0)\;\dot{x}_3^\rho(T_3)  } \nn\\
& & \times \biggl[ \, \Bigl[ 
e^{2i T_2 {\cal F}} \left( e^{2i T_3 {\cal F}} 
- {\bf 1}_L \mbox{Tr}_L\!\left(e^{2i T_3 {\cal F}} \right)
\right) \Bigr]_{\mu\rho} \; e^{S^{(0,-,+)}} \nn \\
& & \;+\;\hspace{1.5pt} \Bigl[  e^{2i T_2 {\cal F}}
\left(  e^{-2i T_1 {\cal F}} - {\bf 1}_L \mbox{Tr}_L
\!\left(e^{-2i T_1 {\cal F}} \right) \right) \Bigr]_{\mu\rho} 
\;e^{S^{(+,-,0)}}\nn \\ 
& & \;+\;\left[ e^{-2i (T_1+T_3) {\cal F}} - {\bf 1}_L 
\mbox{Tr}_L\bigl(e^{-2i (T_1+T_3) {\cal F}} \bigr)
\right]_{\mu\rho}\; e^{S^{(+,0,-)}}\, \biggr]
\quad + \quad ({\cal F}\rightarrow -{\cal F})\ ,
\eeqa
\beqa{gamma2} 
\lefteqn{
\Gamma_2[{\cal A}]  =  -\frac{1}{8}
\int\limits_0^{\infty} 
\!\! dT_1 dT_2 dT_3 \int\!\!d^D\!y_1  \int\!\!d^D\!y_2 \;
\Biggl[  \prod_{k=1}^3 \int\limits_{x_k(0)=y_1}^{x_k(T_k)=y_2}
\!\!\!\!\!\!\!\!\!{\cal D}x_k  \Biggr]
\dot{x}_1^\mu(0)\;\dot{x}_3^\rho(T_3) } \nn\\
& & \;\times\;\biggl[\, \left[ -2i\sin(2{\cal F}(T_2+T_3))
-e^{2i (T_2-T_3) {\cal F}} +
e^{2i T_2 {\cal F}} \mbox{Tr}_Le^{2i T_3 {\cal F}}
\right]_{\mu\rho} \;e^{S^{(0,-,+)}} \nn \\
& & \;+\; \left[ -2i\sin(2{\cal F}(T_1+T_2))
-e^{2i (T_2-T_1) {\cal F}} +
e^{2i T_2 {\cal F}} \mbox{Tr}_Le^{2i T_1 {\cal F}}
\right]_{\mu\rho} \;e^{S^{(+,-,0)}} \nn \\
& & \;+\; \left[ -2\cos(2{\cal F}(T_1+T_3))
+e^{2i (T_3-T_1) {\cal F}} + {\bf 1}_L
\mbox{Tr}_L e^{2i (T_1+T_3) {\cal F}}
\right]_{\mu\rho} \;e^{S^{(+,0,-)}} \,\biggr] \nonumber\\
& & \; + \quad ({\cal F}\rightarrow -{\cal F})\ . 
\eeqa
Then we perform the path integrals of the form
\beq{*0}
<\dot{x}_a^\mu(\tau)\; \dot{x}_b^\nu(\tau') >
_{(\kappa_1,\kappa_2,\kappa_3)}=
\int\!\!d^D\!y_1  \int\!\!d^D\!y_2 \;
\Biggl[  \prod_{k=1}^3 \int\limits_{x_k(0)=y_1}^{x_k(T_k)=y_2}
\!\!\!\!\!\!\!\!\!{\cal D}x_k  \Biggr]
 \dot{x}_a^\mu(\tau)\; \dot{x}_b^\nu(\tau') 
\;e^{S^{(\kappa_1,\kappa_2,\kappa_3)}}  \ ,
\eeq
and this yields
\beq{NG}
<\dot{x}_a^\mu(\tau)\; \dot{x}_b^\nu(\tau') >
_{(\kappa_1,\kappa_2,\kappa_3)} =
{\cal N}^{(\kappa_1,\kappa_2,\kappa_3)} 
\der_\tau \der_{\tau'} {\cal G}_{\mu\nu}^{ab}(\tau,\tau';
\kappa_1,\kappa_2,\kappa_3) \ ,
\eeq
where
\beq{norm3}
{\cal N}^{(\kappa_1,\kappa_2,\kappa_3)} =
(4\pi)^{-D}\,\mbox{det}_L^{-1/2}
\Bigl( \sum_{l=1}^3 
\kappa_l{\cal F}\cot\kappa_l{\cal F}T_l \Bigr) 
\prod_{k=1}^3 T_k^{-D/2}\,\mbox{det}_L^{-1/2}
\left( \frac{\sin \kappa_k {\cal F} T_k}{\kappa_k {\cal F} T_k}
\right)
\int\!\!d^D\!x_0\ ,
\eeq
and~\footnote{In Eq.~\eq{green}, one may replace 
$ G^a_{\mu\nu}(\tau,\tau') \ra G^a_{\mu\nu}(\tau,\tau')-
G^a_{\mu\nu}(\tau,0)- G^a_{\mu\nu}(0,\tau')$ as seen 
in~\cite{WYM}, however our final results do not change because 
we only need the derivatives 
$\der_\tau\der_{\tau'}{\cal G}^{ab}_{\mu\nu}$.}
\beqa{green}
\lefteqn{\hspace{-7mm}
{\cal G}^{ab}_{\mu\nu}(\tau,\tau';\kappa_1,\kappa_2,\kappa_3)=
-\delta_{ab} G^a_{\mu\nu}(\tau,\tau') }\nn\\
&&\hspace{-10mm} +2\Bigl(\bigl[\,
\sum_{k=1}^3\kappa_k{\cal F}\cot(\kappa_k{\cal F}T_k)\,\bigr]^{-1}
\Bigr)_{\rho\sigma}
\Bigl({e^{2i\kappa_a{\cal F}\tau}-1\over e^{2i\kappa_a{\cal F}T_a}-1}
-{1\over2} \Bigr)_{\mu\rho}
\Bigl({e^{2i\kappa_b{\cal F}\tau'}-1\over e^{2i\kappa_b{\cal F}T_b}-1}
-{1\over2} \Bigr)_{\nu\sigma},
\eeqa
with
\beq{*2}
G^a_{\mu\nu}(\tau,\tau') = \left\{ \begin{array}{ll}
{\displaystyle
\delta_{\mu\nu}G^a_B(\tau,\tau') = \delta_{\mu\nu}
\Bigl[\,|\tau-\tau'|-{(\tau-\tau')^2\over T_a}\,\Bigr] }
&\quad (\kappa_a=0) \\ 
{\displaystyle \Bigl[
{1\over2{\cal F}^2}\Bigl({{\cal F}\over\sin({\cal F}T_a)}
e^{-i\kappa_a{\cal F}T_a\der_\tau G^a_B(\tau,\tau')}
+i\kappa_a{\cal F}\der_\tau G^a_B(\tau,\tau')
-{1\over T_a}\Bigr)\Bigr]_{\mu\nu}  }
&\quad (\kappa_a\not=0)\ .  
\end{array}\right.
\eeq
Inserting each value of \eq{NG} 
at $(\tau,\tau')=(0,T_3)$ into Eqs.~\eq{gamma1} and \eq{gamma2}, 
we therefore obtain (the details are presented in Appendix~\ref{ap2})
\beqa{I1int}
\lefteqn{
I_1[{\cal A}]  = -\frac{1}{2} \,(4\pi)^{-D}
\int\limits_0^{\infty} \!\! dT_1 dT_2 dT_3\; \mbox{det}^{1/2}_L 
\left(\frac{{\cal F}^2}{\Delta_{\cal F}}\right) \Biggl\{ } \nn\\
& & \mbox{Tr}_L \Biggl(
  \frac{{\cal F}^2 T_3}{\Delta_{\cal F} \sin{\cal F}T_2}\,
\Bigl[\, 2\sin{\cal F} T_1 \cos 2{\cal F}(T_1+ 2T_2)
  -2\sin{\cal F} (T_1+T_2)\cos{\cal F}(2T_1+3T_2)\nn \\
& & \hspace{10mm}
 +\, \{\,1-2\cos 2{\cal F}(T_1+T_2)\,\}
\sin{\cal F} T_2 \cos{\cal F}(T_1-T_2) \,\Bigr] \nn \\
& & \hspace{10mm}+\,\frac{{\cal F}}{\Delta_{\cal F}}\,\Bigl[\,
4\sin{\cal F}T_1 \sin{\cal F}T_2 \sin 2{\cal F}(T_1+T_2)
-2\sin{\cal F}T_1 \cos{\cal F}(2T_1+3T_2) \nn\\
& & \hspace{10mm} -\,2\sin{\cal F}T_2\cos{\cal F}(T_1-2T_2)
-\sin{\cal F}(T_1+T_2)\cos2{\cal F}(T_1-T_2)\,\Bigr]\,\Biggr)\nn\\
& & \!\!\!\!\!
+\,\mbox{Tr}_L \Biggl( \frac{{\cal F}^2 T_3}{\Delta_{\cal F} 
\sin{\cal F}T_2}\,\Bigl[\, \sin{\cal F}(T_1+T_2) 
\cos{\cal F}(2T_1+T_2) 
-\sin{\cal F}T_1\cos2{\cal F}(T_1+T_2)\,\Bigr] \nn \\
& &\hspace{10mm}
+\,\frac{{\cal F}}{\Delta_{\cal F}}
  \Bigl[ 3\sin{\cal F}T_1 \cos{\cal F}(2T_1+T_2)
+\cos 2{\cal F}T_1 \sin{\cal F}(T_1+T_2)\Bigr]\Biggr) 
\cdot\mbox{Tr}_L \Biggl(\cos 2{\cal F}T_2\Biggr) \nn\\
& & \!\!\!\!\!
+\,\mbox{Tr}_L \Biggl(
\frac{{\cal F}^2 T_3}{\Delta_{\cal F} \sin{\cal F}T_2}\,
\Bigl[\, \sin{\cal F}T_2 \cos{\cal F}(T_1-T_2)
      +\cos{\cal F}T_2 \sin{\cal F}(T_1+T_2) \nn\\
& &\hspace{10mm} -\,\sin{\cal F}T_1\cos 2 {\cal F}T_2\,\Bigr]
+\frac{{\cal F}}{\Delta_{\cal F}}\,\sin{\cal F}T_1\cos{\cal F}T_2
  \Biggr)\cdot\mbox{Tr}_L \Biggl(\cos 2{\cal F}(T_1+T_2)\Biggr) \nn\\
& & +\,\delta (T_2)\,2 (1-D)\,\mbox{Tr}_L 
\Biggl(\cos 2{\cal F}T_1\Biggr) 
+\,\delta (T_3)
\mbox{Tr}_L \Biggl(\cos2{\cal F}(T_1-T_2)\Biggr)\nn\\
&&-\delta(T_3)\mbox{Tr}_L \Biggl(\cos 2{\cal F}T_1\Biggr)\cdot
 \mbox{Tr}_L \Biggl(\cos 2{\cal F}T_2\Biggr) \, \Biggr\}
 \int\!\!d^D\!x_0\ ,
\eeqa
where
\beq{*3}
  \Delta_{\cal F} = \sin{\cal F}T_1\sin{\cal F}T_2
               + {\cal F}T_3\sin{\cal F}(T_1+T_2)\ .
\eeq

\subsection{The $I_2[A]$ part}\label{sec42}
\indent

The computation of the other quantity $I_2[A]$ is similar to the 
above calculations, however the topology of the  world-line 
diagram is different in this case. Let us start with the following 
expression. {}~First, Eq.~\eq{I2} with \eq{Mpseudo} inserted becomes
\beqa{sti2} 
I_2[{\cal A}]  & = & \frac{1}{4} \int\limits_0^{\infty}\!dT_1 dT_2
  \oint [{\cal D}x_1]_{T_1} \oint [{\cal D}x_2]_{T_2}\;
  \delta\!\left(x_1(0)-x_2(0)\right) \nn\\
& &\times\mbox{Tr}_C\biggl[ \lambda^a \! \exp\left\{
{\cal M}^{(+)}_1 \!\otimes{\cal T}_- \right\}_{\mu\nu}\biggr]
\mbox{Tr}_C\biggl[\lambda^a \! \exp\left\{
{\cal M}^{(+)}_2 \!\otimes{\cal T}_- \right\}_{\nu\mu}\biggr]\ .
\eeqa 
With the expansion \eq{expM} and the properties
\beq{*5}
\mbox{Tr}_C \left(\lambda^a\,{\cal T}_{+}\right) = 
\mbox{Tr}_C \left(\lambda^a\,{\cal I}\right)=0\ ,
\eeq
\beq{*6}
\mbox{Tr}_C \left(\lambda^a\,{\cal T}_{-}\right) = 
  n^b\,\mbox{Tr}_C \left(\lambda^a\lambda^b\right) =
  n^b\, 2\,\delta^{ab} = 2\, n^a \ ,
\eeq
we have the formula 
\beq{*7}
\mbox{Tr}_C\biggl[ \lambda^a \! \exp\left\{
  {\cal M}^{(\kappa)}_k \!\otimes
{\cal T}_- \right\}_{\rho\sigma} \biggr] = 
2\,n^a \sinh\left\{ {\cal M}^{(\kappa)}_k \right\}_{\rho\sigma}.
\eeq
Remembering the relation $n^a n^a = 1$, we then derive from~\eq{sti2}
\beqa{*8} 
I_2[{\cal A}] &=& \int\limits_0^{\infty}\!dT_1 dT_2
 \oint [{\cal D}x_1]_{T_1}\oint [{\cal D}x_2]_{T_2}\;
 \delta\!\left(x_1(0)-x_2(0)\right)\mbox{tr}_L\biggl[
  \sinh\left\{  {\cal M}^{(+)}_1 \right\}
  \sinh\left\{  {\cal M}^{(+)}_2 \right\}\biggr] \nn \\
& = & \frac{1}{4}\int\limits_0^{\infty} \!\! dT_1 dT_2
 \biggl[\, {\cal N}^{(+,+)} \mbox{Tr}_L \bigl(
e^{2i(T_1+T_2){\cal F}}\bigr)\,
-  {\cal N}^{(+,-)} \mbox{Tr}_L\bigl(
e^{2i(T_1-T_2){\cal F}}\bigr)\, \biggr]  \nn\\
& & + \quad({\cal F}\rightarrow -{\cal F})\ ,
\eeqa
where
\beq{norm2}
{\cal N}^{(\kappa_1,\kappa_2)} =\int\!\!d^D\!y_1 \int d^Dy_2\;
\Biggl[ \prod_{k=1}^2 \int\limits_{x_k(0)=y_1}^{x_k(T_k)=y_2}
\!\!{\cal D}x_k\Biggr]\delta^D(y_1-y_2)e^{S(\kappa_1,\kappa_2)}
\eeq
with
\beq{action2}
S^{(\kappa_1,\,\kappa_2)} =
  -\frac{1}{4}\sum_{k=1}^2 \int\limits_0^{T_k}\!\!d\tau
  \bigl[\dot{x}_k^2 + 2i\,\kappa_k\,  x_{k\sigma}
{\cal F}_{\sigma\rho}\,\dot{x}_{k\rho} \bigr] \ .
\eeq
The normalizations ${\cal N}^{(\kappa_1,\kappa_2)}$ satisfy the 
following properties:
\beq{*9}
{\cal N}^{(+,+)}={\cal N}^{(+,-)}={\cal N}^{(-,+)}=
{\cal N}^{(-,-)},
\eeq
since, in the present case,  the inversions of paths 
(i.e., the changes of $\kappa$'s signs ) do not change the value of 
the path integral \eq{norm2}: 
note that all the initial and ending points of 
two closed loops are identical. The $({\cal F}\ra-{\cal F})$ terms 
in \eq{*8} lead again to a factor of two,
and thus we have 
\beq{/0} 
I_2[{\cal A}]  =  \frac{1}{2} \int\limits_0^{\infty} 
\!\! dT_1 dT_2\, {\cal N}^{(+,-)} \, \mbox{Tr}_L \bigl(  
e^{2i(T_1+T_2){\cal F}} - e^{2i(T_1-T_2){\cal F}} \bigr)\ .
\eeq
The quantity \eq{norm2} can be evaluated as follows. Recalling the 
relation 
\beq{/1}
\delta^D(y_1-y_2)=\lim_{T_3\ra0}
\left({1\over4\pi T_3}\right)^{D\over2}
\exp\Bigl[\, -{1\over4T_3}(y_1-y_2)^2 \,\Bigr] \ ,
\eeq
we convert the $\delta$-function to the following path integral form:
\beq{/2}
\delta^D(y_1-y_2)=\lim_{T_3\ra0}
\int\limits_{x_3(0)=y_1}^{x_3(T_3)=y_2}
\!\!\!{\cal D}x_3 \,\exp\Bigl[\,
-{1\over4}\int\limits_0^{T_3}{\dot x}_3^2(\tau)d\tau\,\Bigr]\ ,
\eeq 
and this leads to the relations
\beq{Nlim}
{\cal N}^{(\kappa_1,\kappa_2)} =
\lim_{T_3\ra0}{\cal N}^{(\kappa_1,\kappa_2,0)} =
(4\pi)^{-D}\mbox{det}_L^{-1/2}\left(
{\sin{\cal F}T_1 \sin{\cal F}T_2 \over {\cal F}^2}\right)
\int\!\!d^D\!x_0\ .
\eeq
We therefore have the expression
\beqa{I2int}
I_2[{\cal A}]&=&
{1\over2}(4\pi)^{-D}\int\limits_0^\infty dT_1dT_2\,
\mbox{det}_L^{-1/2}\left(
{\sin{\cal F}T_1 \sin{\cal F}T_2 \over {\cal F}^2}\right)\nn\\
&&\times \Bigl[\,\mbox{Tr}_L\cos2{\cal F}(T_1+T_2)-
\mbox{Tr}_L\cos2{\cal F}(T_1-T_2)\,\Bigr] 
\int\!\!d^D\!x_0\ .
\eeqa
It is interesting that the normalization 
${\cal N}^{(\kappa_1,\kappa_2)}$ can still be obtained as 
a singular limit (the $T_3\ra0$ limit) from Eq.~\eq{norm3}, although 
$I_2$ is not a singular part of $I_1$ (cf. \eq{I1int}) ~\cite{WYM}. 

\section{The world-line moduli integrals}\label{sec5}
\setcounter{section}{5}
\setcounter{equation}{0}
\indent

In this section, we study the divergence structure of the 
Euler-Heisenberg-type action derived in the previous sections. 
In the world-line formalism, divergences come out explicitly after 
performing the proper time integrals (the $T_a$ integrals; $a=1,2,3$ 
in the present case), and hence we have to examine how to perform 
these integrals before discussing renormalizations. However, generally 
speaking, multi-integrations are difficult to perform in an analytic 
way, and hence we here consider the simplest case corresponding to 
gluon two-point function parts. 

Let us consider the Taylor expansions of $\Gamma[A]$ concerning 
${\cal F}$ (omitting constant terms) in the same way as done in 
the one-loop case. Expanding $I_1$ and $I_2$ up to the second order 
of ${\cal F}$, we extract the following quantities from 
Eqs.~\eq{I1int} and \eq{I2int}: 
\beqa{I1Ci,I2Ci}
I_1[{\cal A}]&=& 
{g_0^4\mu^{4\eps} \over(4\pi)^{4-2\eps}} 
\int\!\!d^D\!x_0\,
{\cal F}_{\mu\nu}{\cal F}_{\nu\mu}\Bigl\{\,
30 C_1 +42 C_2 + 48C_3 +{93\over2}C_4-2C_5-10C_6\nn\\
&&-\,(17C_1+29C_2+{57\over2}C_3+32C_4-{17\over3}C_6)\,\eps\nn\\
&&-\,(2C_1+2C_2+5C_3+6C_4-{2\over3}C_6)\,\eps^2 \,\Bigr\} 
+\cdots\ , \label{I1Ci}\\
I_2[{\cal A}]&=&{g_0^4\mu^{4\eps} \over(4\pi)^{4-2\eps}} 
\int\!\!d^D\!x_0\,
{\cal F}_{\mu\nu}{\cal F}_{\nu\mu}\, (-4)\,C_5 +\cdots\ ,
\label{I2Ci}
\eeqa
with
\beq{Ci}
\begin{array}{lcl}
C_1 &=& \int\limits_0^\infty\,dT_1dT_2dT_3\,\Delta^{\eps-4}\,
        T_1^4T_2\, e^{-m^2(T_1+T_2+T_3)} \\
C_2 &=& \int\limits_0^\infty\,dT_1dT_2dT_3\,\Delta^{\eps-4}\,
        T_1^3T_2^2\, e^{-m^2(T_1+T_2+T_3)} \\
C_3 &=& \int\limits_0^\infty\,dT_1dT_2dT_3\,\Delta^{\eps-4}\,
        T_1^3T_2T_3\, e^{-m^2(T_1+T_2+T_3)} \\
C_4 &=& \int\limits_0^\infty\,dT_1dT_2dT_3\,\Delta^{\eps-4}\,
        T_1^2T_2^2T_3\, e^{-m^2(T_1+T_2+T_3)} \\
C_5 &=& \int\limits_0^\infty\,dT_1dT_2dT_3\,\Delta^{\eps-4}\,
        T_1^3T_2^3\, \delta(T_3)\,e^{-m^2(T_1+T_2+T_3)} \\
C_6 &=& \int\limits_0^\infty\,dT_1dT_2dT_3\,\Delta^{\eps-4}\,
        T_1^4T_2^2\, \delta(T_3)\,e^{-m^2(T_1+T_2+T_3)}\ , 
\end{array}
\eeq
where we have put $D=4-2\eps$ and $\Delta=T_1T_2+T_2T_3+T_3T_1$.
Also, the damping mass factor $e^{-m^2(T_1+T_2+T_3)}$ is inserted in 
each $C_i$ in the same way as in  the one-loop calculation 
(Section~\ref{sec3}). In the meantime, we shall introduce the 
notation $\eps'=\{\eps;\eps>1\}$ for $C_1$ and $C_6$, in order not 
to mix it up with the (usual) infinitesimal parameter $\eps>0$. 
This description is indispensable for the convergency of the 
$C_1$ and $C_6$ integrals, although we shall set $\eps'=\eps$ after 
all, expecting the analytic continuation (see also Section 8-1-2 
in \cite{IZ}). The $\eps'$ divergences are related to the divergences 
from the artificial mass term insertion, since $C_1$ and $C_6$ 
contain tadpole contributions, which vanish in the $m\ra0$ limit 
(in the sense of dimensional regularization).

In $C_5$ and $C_6$, all the integrals are easy to perform, and hence 
we simply write down
\beqa{C56}
C_5 & =& (m^2)^{-2\eps}\, \Gamma^2(\eps)\ , \\ 
C_6 & =& (m^2)^{-2\eps}\, \Gamma(\eps+1)\Gamma(\eps'-1)\ .
\eeqa
The rest of $C_i$ are computed in detail in Appendix~\ref{ap3}, 
and we only show the results as follows:
\beqa{CiF32}
C_1 &=& (m^2)^{-2\eps} {\Gamma(2\eps)\over3-\eps}
\mbox{B}(\eps'-1,\eps+2) + (m^2)^{-2\eps}{1\over64}
\Gamma(\eps+3)\Gamma(\eps-3) \nn\\
& &\times\Bigl[\, 4\mbox{B}(2,{1\over2})
{}~F(2,3+\eps,{5\over2},{1\over4}) - 3\mbox{B}(3,{1\over2})
{}~F(3,3+\eps,{7\over2},{1\over4})\,\Bigr] \nn\\
&+& (m^2)^{-2\eps} 4^{-\eps}{\Gamma(2\eps+1)\over(3-\eps)(\eps-2)}
\Bigl[\, \mbox{B}(\eps,{1\over2})
{}_3F_2(1,2\eps+1,\eps;\eps-1,\eps+{1\over2};{1\over4})\nn\\ 
& &- {3\over4} \mbox{B}(\eps+1,{1\over2}) {}_3F_2(1,2\eps+1,
\eps+1;\eps-1,\eps+{3\over2},{1\over4})\,\Bigr] \ ,
\label{C1F32}\\
C_2 &=&  (m^2)^{-2\eps} {1\over64} \Gamma(3+\eps) 
\Gamma(\eps-3) \mbox{B}(3,{1\over2})
{}~F(3,3+\eps;{7\over2};{1\over4}) \nn\\
&+&  (m^2)^{-2\eps} 4^{-\eps} {\Gamma(2\eps)\over3-\eps}
 \mbox{B}(\eps,{1\over2}){}_3F_2(2\eps,1,\eps;\eps-2,
\eps+{1\over2};{1\over4})\ , \label{C2F32}\\
C_3 &=& - C_4 + (m^2)^{-2\eps} {1\over16}
{\Gamma(4+\eps)\Gamma(\eps-2)\over(\eps+2)(\eps+3)}
\mbox{B}(2,{1\over2})
{}_3F_2(4-\eps,2+\eps,2;3-\eps,{5\over2};{1\over4})\nn\\
&+&(m^2)^{-2\eps} 4^{-\eps} \Gamma(2\eps)
{\Gamma(2-\eps)\over\Gamma(4-\eps)}\mbox{B}(\eps,{1\over2}) 
{}_3F_2(2,,2\eps,\eps;\eps-1,\eps+{1\over2};{1\over4})\ ,
\label{C3F32} \\
C_4&=& (m^2)^{-2\eps}{1\over32} {\Gamma(4+\eps)
\Gamma(\eps-2)\over(\eps+2)(\eps+3)} \mbox{B}(3,{1\over2})
{}_3F_2(4-\eps,2+\eps,3;3-\eps,{7\over2};{1\over4}) \nn\\
&+& (m^2)^{-2\eps}{1\over2}4^{-\eps} \Gamma(2\eps) 
{\Gamma(2-\eps)\over\Gamma(4-\eps)}\mbox{B}(\eps+1,{1\over2}) 
{}_3F_2(2,2\eps,\eps+1;\eps-1,\eps+{3\over2};{1\over4})\ . 
\label{C4F32}
\eeqa

We can rewrite these expressions in terms of the hypergeometric 
function $F\equiv{}_2F_1$  only, and the results and their 
derivations are presented in Appendix~\ref{ap3} (see \eq{C1F21}, 
\eq{C2F21}, \eq{C3F21} and \eq{C4F21}). Since we could not find 
any convenient transformation formula from ${}_3F_2$ to ${}_2F_1$ 
in the literature,  we have established a transformation technique 
in Appendix~\ref{ap3}. (A more concise explanation can be found 
in Appendix~\ref{ap4}.)  

Let us consider the expressions \eq{C1F21}, \eq{C2F21}, \eq{C3F21} 
and \eq{C4F21}. We now perform the Taylor expansions of the 
hypergeometric functions around $\eps=0$, in order to see the 
divergence structures of the coefficients $C_i$. Here, we are 
only interested in $1/\eps$ terms in the sense of the MS scheme, 
and the hypergeometric functions which contribute to the desired 
pole terms are only generated from the following expansion:
\beqa{taylor}
\lefteqn{F(\alpha,a\eps;\gamma+b\eps;{1\over4})
=F(\alpha,0;\gamma;{1\over4}) }\nn\\
&&+\,a\eps F^{(0,1,0)}(\alpha,0;\gamma;{1\over4})
+\,b\eps F^{(0,0,1)}(\alpha,0;\gamma;{1\over4})
+{\cal O}(\eps^2)\ ,
\eeqa
where 
\beq{/4}
{}~F^{(n,m,l)}(\alpha,\beta;\gamma;z)
\define \der^n_\alpha\,\der^m_\beta\,\der^l_\gamma\,
{}F(\alpha,\beta;\gamma;z) \ .
\eeq
These differential coefficients (for $\gamma\not=0$) are evaluated by 
\beqa{Form2,3}
& &F(\alpha,0;\gamma;z)=1\ ,\qquad
F^{(0,0,1)}(\alpha,0;\gamma;z)=0\ , \label{Form2}\\
& &F^{(0,1,0)}(\alpha,0;\gamma;z)=
F^{(1,0,0)}(0,\alpha;\gamma;z)= {\alpha\over\gamma}z\,
{}_3F_2(1,1,\alpha+1;2,\gamma+1;z)\ .\label{Form3}
\eeqa
Here we again encounter the generalized hypergeometric function 
${}_3F_2$, however in the present case, it can be reduced to 
the ordinary hypergeometric function ${}_2F_1$ through 
the following formula (derived in Appendix~\ref{ap4}):
\beq{Form4}
{}_3F_2(1,\beta,n+1;2,\gamma;z)
= {1\over n}\sum_{k=1}^n F(k,\beta;\gamma;z),
\qquad (n\geq1,|z|<1, \Re(\gamma) >0).
\eeq
Combining \eq{Form3} and \eq{Form4}, we have 
\beq{Form5}
F^{(0,1,0)}(n,0;\gamma;z)=
{z\over\gamma}\sum_{k=1}^n F(k,1;\gamma+1;z)\ ,
\qquad (n\geq1,|z|<1, \Re(\gamma) >0), 
\eeq
and thus
\begin{equation}
  F(n,a\eps;\gamma+b\eps;z) =
  1 + \eps\frac{az}{\gamma}\sum_{k=1}^n F(k,1;\gamma+1;z)
  + {\cal O}(\eps^2)\ ,
  \qquad (n\geq1,|z|<1, \Re(\gamma) >0). 
\end{equation}
Owing to this formula, all coefficients in front of $1/\eps$ 
can be written in terms of ${}_2F_1$ and the gamma functions. 
After some algebra, we obtain
\beqa{Cipole}
&&C'_1=-{1\over6\eps^2}+\Bigl(-{5\over9}+{\rho_m\over3}\Bigr)
{1\over\eps}+{\cal O}(1)\ , \qquad
C'_2={1\over6\eps^2}+\Bigl({1\over18}-{\rho_m\over3}\Bigr)
{1\over\eps}+{\cal O}(1)\ , \nn\\
&&C'_3={1\over12\eps^2}-\Bigl({1\over72}+{\rho_m\over6}\Bigr)
{1\over\eps}+{\cal O}(1)\ , \qquad
C'_4=-{1\over12\eps}+{\cal O}(1)\ , \nn\\
&&C'_5={1\over\eps^2}-{2\rho_m\over\eps}+{\cal O}(1)\ ,\qquad
C'_6=-{1\over\eps}+{\cal O}(1)\ ,
\eeqa
where the overall factor $(4\pi\mu^2)^{2\eps}$ seen in 
\eq{I1Ci} and \eq{I2Ci} is absorbed in $C_i$; i.e., 
\beq{newCi}
C'_i = (4\pi\mu^2)^{2\eps}C_i\ ,
\eeq
and we have defined
\beq{rhom}
\rho_m=\gamma_{\scriptscriptstyle{E}}+\ln{m^2\over4\pi\mu^2}\ ,
\qquad \gamma_{\scriptscriptstyle{E}}=\mbox{Euler const.}\ 
\eeq
Therefore from Eqs.~\eq{I1Ci}, \eq{I2Ci} and \eq{Cipole}, we obtain
\begin{eqnarray}
  I_1[{\cal A}] & = &
   \frac{4g_0^4}{(4\pi)^4}
   \left[
     \frac{4}{\eps^2} + \left(-\frac{11}{2} -8\rho_m\right)\frac{1}{\eps}
   \right] 
   \int\!\!d^D\!x_0\left(-\frac{1}{4}{\cal F}_{\mu\nu}{\cal F}_{\mu\nu}
   \right) \quad+\quad {\cal O}(\eps^0) \\
  I_2[{\cal A}] & = & 
   \frac{4g_0^4}{(4\pi)^4}
   \left[
     -\frac{4}{\eps^2} + \frac{8\rho_m}{\eps}
   \right] 
   \int\!\!d^D\!x_0\left(-\frac{1}{4}{\cal F}_{\mu\nu}{\cal F}_{\mu\nu}
   \right) \quad+\quad {\cal O}(\eps^0) 
\end{eqnarray}
and due to Eq.~\eq{ga}
the renormalization part of the effective action (purely gluon parts) 
at the 2nd order in ${\cal F}$ in our regularization is found to be 
\beq{finalGA}
\Gamma[{\cal A}]= 
   \frac{4g_0^4}{(4\pi)^4}
   \left(
     -\frac{11}{2\eps} 
   \right) 
   \int\!\!d^D\!x_0\left(-\frac{1}{4}{\cal F}_{\mu\nu}{\cal F}_{\mu\nu}
   \right) \quad+\quad {\cal O}(\eps^0) 
\eeq

%

\vspace{8mm}
\begin{minipage}[htb]{16cm}
\begin{center}
\epsfxsize=14cm
\unitlength=1cm
\begin{picture}(16,4)
\put(0,0.6){\epsfbox{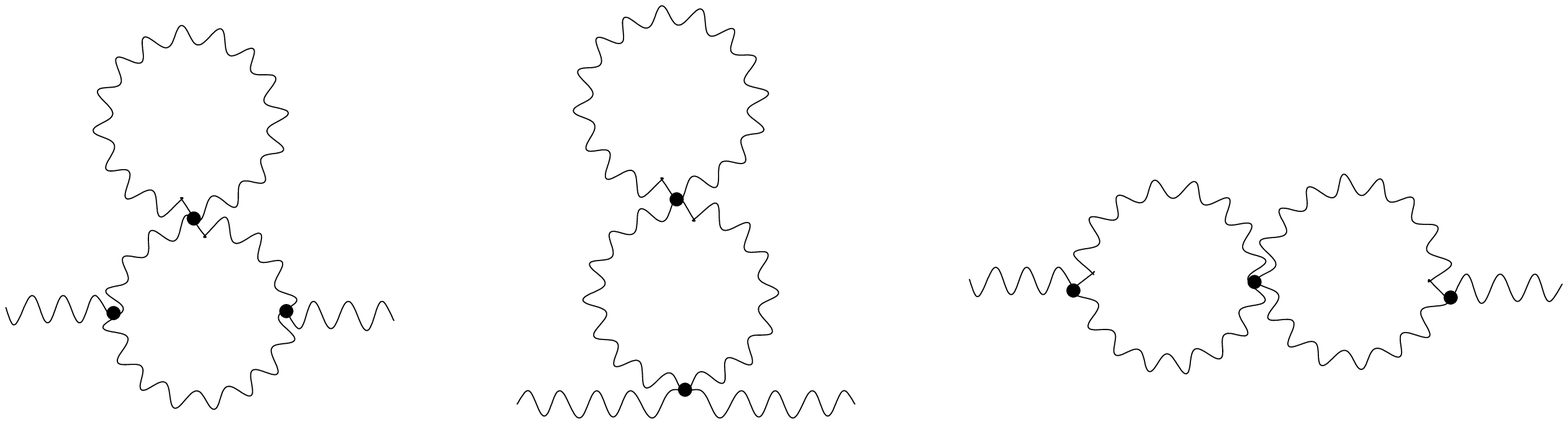}}
\put(1.5,0){$\mbox{(a)}$}
\put(5.9,0){$\mbox{(b)}$}
\put(11,0){$\mbox{(c)}$}
\end{picture}
\end{center}
{\bf Figure 2:} The ``eight-figure'' diagrams contained in the 
coefficients $C_5$ and $C_6$.
\end{minipage}
\vspace{5mm}

{}~Finally, let us put a comment on what our results imply. 
Picking up $C_5$ and $C_6$ from Eqs.~\eq{I1Ci} and \eq{I2Ci}, 
and using \eq{FT}, we extract the following quantities corresponding 
to self-energy parts:
\beqa{pi5,6}
&&{\Pi_5}^{ab}_{\mu\nu}=
g_0^4{4\delta^{ab}\over(4\pi)^4} 
(k^2\delta_{\mu\nu}-k_\mu k_\nu)(-6 C_5')\ ,\label{pi5}\\
&&{\Pi_6}^{ab}_{\mu\nu}=
g_0^4{4\delta^{ab}\over(4\pi)^4} 
(k^2\delta_{\mu\nu}-k_\mu k_\nu)(-10 C_6')\ .\label{pi6}
\eeqa
As briefly shown in Appendix~\ref{ap5}, the $\Pi_5$ and $\Pi_6$ 
exactly coincide with the Feynman diagram results, if the 
coefficients of gluon kinetic terms are evaluated in the region 
close to the light cone $k^2\ra0$; in other words, if  $k^2$ is much 
smaller than the mass parameter $m^2$. (Note that 
$C_A=2$ in the $su(2)$ case.) Thus, it is very natural to expect 
that the other coefficients $C_1$, $C_2$, $C_3$ and $C_4$ should 
possess the same meaning.  

\section{Conclusions and Discussions}\label{sec6}
\setcounter{section}{6}
\setcounter{equation}{0}
\indent

In this paper, we have explicitly calculated the gluon parts of the 
two-loop Euler-Heisenberg actions, which are organized at the level 
of an implicit formulation in the previous paper~\cite{WYM}. The 
present results are still preliminary to reach a clear physical 
quantity such as $\beta$-functions, however this paper is an 
important step toward a full-fledged extension of the world-line 
formalism to two-loop Yang-Mills theories. One of the main obstacles 
for this aim is the problem of how to integrate the proper time 
variables at higher loop calculations. Also, in the sense of field 
theory limit of string theory, this is an important problem: 
at the level of string theory, it is recognized as the problem how 
to perform the moduli integrals on a multiloop world-sheet. 

We have performed the path integrals in Section~\ref{sec4}, 
applying the world-line Green functions to all combinations of 
$su(2)$-like charges (the $\kappa$ signatures; q.v. \eq{action3},
\eq{Wform}, \eq{gamma1} and \eq{gamma2}). Then extracting the parts 
corresponding to the wave function renormalization, we have 
been able to perform the world-line ``moduli'' integrals to reveal 
the divergence structure in Section~\ref{sec5}. In the standard 
method, it is difficult to perform the loop integrals containing a 
mass parameter, while in our case, all the integrations are carried 
out in Appendix~\ref{ap3}, giving rise to (generalized) hypergeometric 
functions as a result. This is certainly a significant point from a 
theoretical viewpoint, and hence we have mainly focused on the 
technical issue concerning the integrations. We should also note 
that the pseudo-abelian technique has worked out both at one- and 
two-loop levels, with reproducing the Feynman diagram results of 
Appendix~\ref{ap5}. We expect that our integration method will 
straightforwardly apply to higher order terms in ${\cal F}$ as well. 
 
Although we have verified that our results contain correct Feynman 
diagram contributions, the followings should further be investigated 
as a next step toward our goal: the coincidence between the present 
results and those by the Feynman diagram method can only be understood 
in the region close to the light cone $k^2\ra0$ ($k^2<<m^2$), where 
$k_\mu$ are the external gluon momenta. On the other hand, as seen 
in Section~\ref{sec3}, we do not have the restriction on $k^2$ at 
the one-loop level, in order to extract the $\beta$-function. 
Similarly we shall encounter the same difference in the ghost loop 
calculations, and should clarify the reason for this kind of 
discrepancy. Related to this issue, another question is whether or 
not we can evaluate the pole structures of $C_i$ for $m=0$ (or 
$m^2<<k^2$). As inferred from Eqs.~\eq{limPT} and \eq{limPe}, the 
integrals $C_i$ might depend on the region of either $k^2<<m^2$ or 
not. However the present calculations do not indicate such a 
dependence, simply because $C_i$ are not the Fourier modes of 
gluon two-point function. To clarify this point, one should compute 
the correlator of two gluon vertex operators (of bosonic field 
representation) along the outline of Appendix B in Ref.~\cite{WYM}; 
in this case, the formulation should be extended to the 
super world-line formalism in order to optimize the inclusion of 
the four point interactions involving external legs. 
Anyway, in order to find the correct $\beta$-function coefficient 
at two loops, we also have to add the contributions including the 
counter terms generated from one-loop divergences (in the massive 
formulation). 

Following ref.~\cite{CMM} a gauge symmetry breaking IR gluon 
mass $m^2$ (introduced there only as a device to seperate IR and UV
divergences) requires a further counter term
\begin{equation}
  \label{eq:counter}
  \frac{1}{2}Z_x m^2 A_{\mu}^aA_{\mu}^a
  \qquad \mbox{with}\qquad Z_x =-\frac{g^2 C_A}{16\pi^2\varepsilon} 
\end{equation}
and  $C_A=2$ in our case, cancelling $m^2$ dependent singularities in 
the MS dimensional regularization scheme. 
Insertion of this counter term `gluon mass' into the one-loop contribution of
order ${\cal F}^2$ 
as found in Eq.~(\ref{gf2}), i.e.~calculating with 
$m^2 \rightarrow m^2 + Z_x m^2$ and expanding to first
order in $\frac{m^2}{\varepsilon}$, thus using
\begin{equation}
e^{-m^2 T} \rightarrow e^{-m^2 T}\biggl(1-\frac{g^2 m^2 C_A}
{16\pi^2\varepsilon}T \biggr)  \ ,
\end{equation}
yields, e.g.~a singular two-loop 
${\cal F}^2$-contribution:
\begin{eqnarray} \label{Gct}
\Gamma_{ct} & = &  
\frac{C_A}{2(4\pi)^{D/2}}
\Bigl({D\over12}-2\Bigr)
\int\limits_0^\infty\!\! dT\,T^{1-D/2}\,
e^{-m^2T} 
\Bigl(
  \frac{-g^2 m^2 C_A}{16\pi^2\varepsilon}T
\Bigr)
\int\!\!d^D\!x_0\,\mbox{Tr}_L{\cal F}^2 \\
& = &
\frac{g^4 C_A^2}{(4\pi)^4}
\biggl({10\over3\eps}\biggr)
\int\!\!d^D\!x_0\,
\Biggl(-\frac{1}{4}{\cal F}_{\mu\nu}{\cal F}_{\mu\nu} \Biggr)
+{\cal O}(\eps^0)\ .
\end{eqnarray}
However in the background formalism there should be further counter terms
including quantum as well as background fields. This has to be further
analyzed in order to reproduce the usual $\beta$-function coefficient.

This paper concerns a theoretical interest, and is an important 
part of the ongoing long-term effort to find a way to getting over 
difficulties in the current calculation methods for higher loop 
amplitudes in field theory. Since the whole computation process in 
the world-line formalism looks completely different from the standard 
field theory calculations, the above questions are the milestones 
in the future study, and should be solved in order to make future 
practical applications successful.


\section*{Acknowledgement}

We would like to thank M.~Jamin for helpful discussions concerning
infrared regularization.

\begin{appendix}
\section{List of path reversal formulae}\label{ap1}
\setcounter{section}{1}
\setcounter{equation}{0}
\indent

This appendix is a brief note on the reversals of path ordering 
and proper time directions. The standard definition of the 
path ordering (the normal type) is 
\beq{/5}
  \mbox{P} \exp\left\{
  \int\limits_{\tau_{\alpha}}^{\tau_{\beta}}d\tau
  M_{\tau}[x] \right\} =
  \sum_{n=0}^{\infty} 
  \int\limits_{\tau_{\alpha}}^{\tau_{\beta}}d\tau_1
  \int\limits_{\tau_{\alpha}}^{\tau_1}d\tau_2\cdots\!\!
  \int\limits_{\tau_{\alpha}}^{\tau_{n-1}}\!\!d\tau_n
  M_{\tau_{1}}[x]\cdots M_{\tau_n}[x]\ ,
\eeq
and the anti-path ordering (used in~\cite{WYM} for a certain reason) 
is
\beq{/6}
  \mbox{P}^* \exp\left\{
  \int\limits_{\tau_{\alpha}}^{\tau_{\beta}}d\tau
  M_{\tau}[x] \right\} =
  \sum_{n=0}^{\infty} 
  \int\limits_{\tau_{\alpha}}^{\tau_{\beta}}d\tau_1
  \int\limits_{\tau_{\alpha}}^{\tau_1}d\tau_2\cdots\!\!
  \int\limits_{\tau_{\alpha}}^{\tau_{n-1}}\!\!d\tau_n
  M_{\tau_{n}}[x]\cdots M_{\tau_1}[x] \ .
\eeq
We here assume that the $M_\tau$ in the above two definitions 
are the same objects. 

The relations between the path and the anti-path ordering formulae 
are given by  
\beq{aW}
W^{ae}_{\mu\nu}[x;T,0]=\mbox{P}\exp\left\{\int\limits_0^T
 d\tau M_{\tau}[x]\right\}_{\mu\nu}^{ae}
=\mbox{P}^*\exp\left\{\int\limits_0^T
 d\tau M^T_{\tau}[x]\right\}_{\nu\mu}^{ea} \ ,
\eeq
\beq{/7}
 W^{ae}_{\mu\sigma\rho\nu}[x;S,\tau_\alpha,0] =
 \mbox{Tr}_C\biggl[
  \lambda^a \,\mbox{P}^*\exp\left\{
  \int\limits_0^{\tau_\alpha}
     d\tau M^T_{\tau}[x]\right\}_{\nu\rho}
  \!\!\!\!\!
  \lambda^e \,\mbox{P}^*\exp\left\{                   
  \int\limits_{\tau_\alpha}^S
     d\tau M^T_{\tau}[x]\right\}_{\sigma\mu}
  \biggr]  \ ,
\eeq
where the $M^T$ represents the transposition on both the color 
and Lorentz spaces. 

Note that the transposition makes an additional minus 
sign in front of the $A{\dot x}$ term in $M_\tau$. 
If we change the sign of gauge coupling $g$, this additional 
sign drops out, and the starting formulae presented in Sect.~\ref{sec2}
follows from the previous paper directly.

The other useful observation is related to path inversion: As can be seen 
from Eq.~(\ref{defM}) by distinguishing $\tau$ and $\tau^{\prime}$ carefully
(see Eq.~(\ref{MT_detail})), the following relation holds:
\begin{equation}
  \label{eq:Mtrans}
  \left(M_{\tau}[x^{-1}]\right)^{ab}_{\mu\nu}
  =
  \left(M_{\tau_{\alpha}+\tau_{\beta}-\tau}[x]\right)^{ba}_{\nu\mu}\ ,
\end{equation}
where $x^{-1}$ denotes the inverted path $x$, each of them defined by
\begin{eqnarray}
  x: [\tau_{\alpha}, \tau_{\beta}] \rightarrow {R}^4,& &
  \quad\tau\mapsto  x(\tau)\ ,\\
  x^{-1}: [\tau_{\alpha}, \tau_{\beta}] \rightarrow {R}^4,& &
  \quad\tau\mapsto x^{-1}(\tau):=x(\tau_{\alpha}+\tau_{\beta}-\tau)
\end{eqnarray}
respectively. As a consequence of Eq.~(\ref{eq:Mtrans}) we have
\beq{/8}
\mbox{P}\exp\left\{\int\limits_{\tau_{\alpha}}^{\tau_{\beta}}
 \!\!d\tau M_{\tau}[x]\right\}^T
 =
\mbox{P}\exp\left\{\int\limits_{\tau_{\alpha}}^{\tau_{\beta}}
 \!\!d\tau M_{\tau}[x^{-1}]\right\}\ ,
\eeq
where again $T$ means transposition in both color as well as Lorentz 
space.

Describing path inversion by the $x^{-1}$ symbol is a useful tool in 
performing world-line calculations (rather than changing the direction of 
$\tau$ or other possibilities), because it nicely fits together with 
the following identities for world-line path integrals\footnote{These 
identities can easily be verified from the path integral discretization.}:
\begin{equation} \label{PFADINV}
  \int\limits_{x(0)=y_1}^{x(S)=y_2}\!\!\!\!\!\!{\cal D}x\;F[x]
  \quad =
  \int\limits_{x(0)=y_2}^{x(S)=y_1}\!\!\!\!\!\!{\cal D}x\;F[x^{-1}]  
\end{equation}
and consequently
\begin{equation} \label{gpfadinv}
  \oint{\cal D}x\;F[x]
  =
  \int\!\!d^D\!y\int\limits_{x(0)=y}^{x(S)=y}\!\!\!\!\!\!{\cal D}x\;F[x]
  =
  \int\!\!d^D\!y\int\limits_{x(0)=y}^{x(S)=y}\!\!\!\!\!\!{\cal D}x\;F[x^{-1}]
  =
  \oint{\cal D}x\;F[x^{-1}]  \ ,
\end{equation}
where $F$ is an arbitrary functional and the integrands $F[x^{-1}]$  
are to be understood as follows: for any path $x$ integrated over within 
the path integral the corresponding path $x^{-1}$ is to be constructed 
(in thoughts) and the functional is to be evaluated for this path $x^{-1}$. 
Thus, in the path integral, $x^{-1}$ depends on $x$. 
Note that we also have
\begin{eqnarray}
  \int\limits_{x(0)=y_1}^{x(S)=y_2}\!\!\!\!\!\![{\cal D}x]_S\;F[x]
  \quad & = &
  \int\limits_{x(0)=y_1}^{x(S)=y_2}\!\!\!\!\!\!{\cal D}x\;
  e^{-\int\limits_0^S\!\! d\tau\;
  \frac{1}{4}\dot{x}^2}F[x]
  =
  \int\limits_{x(0)=y_2}^{x(S)=y_1}\!\!\!\!\!\!{\cal D}x\;
  e^{-\int\limits_0^S\!\! d\tau\;
  \frac{1}{4}(\dot{x}^{-1})^2}F[x^{-1}] \nonumber\\
  & = &
  \int\limits_{x(0)=y_2}^{x(S)=y_1}\!\!\!\!\!\!{\cal D}x\;
  e^{-\int\limits_0^S\!\! d\tau\;
  \frac{1}{4}\dot{x}^2}F[x^{-1}]
  =
  \int\limits_{x(0)=y_2}^{x(S)=y_1}\!\!\!\!\!\![{\cal D}x]_S\;F[x^{-1}]\ ,
\end{eqnarray}
i.e.~our usual bracket notation for the free path integral part is not 
affected.

Using these relations and identities one can easily verify that the
gluon propagator in a backgound, given by the world-line 
representation~\cite{RSS} 
\beq{:}
\Delta_{\mu\nu}^{ab}(x_1,x_2) =\int\limits_0^{\infty} dT \!\!
  \int\limits_{x(0)=x_2}^{x(T)=x_1} \!\![{\cal D}x]_T\;
  \mbox{P}\exp\left\{
  \int\limits_0^T d\tau M_{\tau}[x]\right\}^{ab}_{\mu\nu}\ ,
\eeq
satisfies the following property:
\beq{/9}
\Delta^{ab}_{\mu\nu}(x_1,x_2)=\Delta^{ba}_{\nu\mu}(x_2,x_1)\ .
\eeq

\section{The derivation of $I_1[{\cal A}]$}\label{ap2}
\setcounter{section}{2}
\setcounter{equation}{0}
\indent

In this Appendix, we show some details for the computation in 
Section~\ref{sec41}. {}~First, we need various values of the 
quantity~\eq{NG} at $(\tau,\tau')=(0,T_3)$ for $a,b=1,3$, and 
those are given by the following. The necessary derivatives of the 
Green functions are
\beqa{::} \label{delGs_start}
\der_\tau\der_{\tau'}{\cal G}^{13}_{\mu\nu}(\tau,\tau';+,0,-)
\Bigr|_{\tau=0,\tau'=T_3} \dand=\dand \left(\, {2T_2{\cal F}^2 \,
e^{i{\cal F}(T_3-T_1)} \over \sin{\cal F}T_1\sin{\cal F}T_3
+{\cal F}T_2\sin{\cal F}(T_1+T_3)} \,\right)_{\mu\nu}      \\
\der_\tau\der_{\tau'}{\cal G}^{13}_{\mu\nu}(\tau,\tau';0,-,+)
\Bigr|_{\tau=0,\tau'=T_3} \dand=\dand \left(\, {2{\cal F}\sin{\cal F}T_2 
\,e^{-i{\cal F}T_3} \over \sin{\cal F}T_2\sin{\cal F}T_3
+{\cal F}T_1\sin{\cal F}(T_2+T_3)} \,\right)_{\mu\nu}      \\
\der_\tau\der_{\tau'}{\cal G}^{13}_{\mu\nu}(\tau,\tau';+,-,0)
\Bigr|_{\tau=0,\tau'=T_3} \dand=\dand \left(\, {2{\cal F}\sin{\cal F}T_2
\, e^{-i{\cal F}T_1} \over \sin{\cal F}T_1\sin{\cal F}T_2 \,
+{\cal F}T_3\sin{\cal F}(T_1+T_2)} \,\right)_{\mu\nu}     \\
\der_\tau\der_{\tau'}{\cal G}^{33}_{\mu\nu}(\tau,\tau';+,0,-)
\Bigr|_{\tau=0,\tau'=T_3} \dand=\dand  \left(\, {2T_2{\cal F}^2 \,
e^{i2{\cal F}T_3} \sin{\cal F}T_1\,\mbox{cosec}\,{\cal F}T_3
\over \sin{\cal F}T_1\sin{\cal F}T_3 
+{\cal F}T_2\sin{\cal F}(T_1+T_3)} \,\right)_{\mu\nu} \nn\\
&&+\,2\left(\,{\bf 1}_L\delta(T_3)
-{{\cal F}\over\sin{\cal F}T_3}e^{i{\cal F}T_3}
\,\right)_{\mu\nu}    \\
\der_\tau\der_{\tau'}{\cal G}^{33}_{\mu\nu}(\tau,\tau';0,-,+)
\Bigr|_{\tau=0,\tau'=T_3} \dand=\dand  \left(\, {2T_1{\cal F}^2 \,
e^{-i2{\cal F}T_3} \sin{\cal F}T_2\,\mbox{cosec}\,{\cal F}T_3
\over \sin{\cal F}T_2\sin{\cal F}T_3 
+{\cal F}T_1\sin{\cal F}(T_2+T_3)} \,\right)_{\mu\nu} \nn\\
&&+\,2\left(\,{\bf 1}_L\delta(T_3)
-{{\cal F}\over\sin{\cal F}T_3}e^{-i{\cal F}T_3}
\,\right)_{\mu\nu}    \\ 
\der_\tau\der_{\tau'}{\cal G}^{33}_{\mu\nu}(\tau,\tau';+,-,0)
\Bigr|_{\tau=0,\tau'=T_3} \dand=\dand \left(\, {2 T_3^{-1}
\sin{\cal F}T_1 \sin{\cal F}T_2 \over 
\sin{\cal F}T_1\sin{\cal F}T_2 
+{\cal F}T_3\sin{\cal F}(T_1+T_2)} \,\right)_{\mu\nu}\nn\\
&&+\,2(\delta(T_3)-{1\over T_3})({\bf 1}_L)_{\mu\nu}\ .
\label{delGs_end}
\eeqa
The normalization constants are given by 
\beq{:.}
  {\cal N}^{(+,-,0)} = (4\pi)^{-D} \mbox{det}^{1/2}_L
  \left(\frac{{\cal F}^2}{\sin{\cal F}T_1\sin{\cal F}T_2
   + {\cal F}T_3\sin{\cal F}(T_1+T_2)} \right)
\int\!\!d^D\!x_0\ ,
\eeq
and the other values can be obtained by exchanging the labels $a$ 
on $\kappa_a$ and $T_a$ simultaneously; for example, 
\beq{.:} 
{\cal N}^{(+,0,-)} ={\cal N}^{(+,-,0)}\Bigr|_{T_2\leftrightarrow T_3}.
\eeq
Note also
\beq{:*}
{\cal N}^{(+,+,0)} ={\cal N}^{(-,-,0)}={\cal N}^{(\pm,\mp,0)}\ .
\eeq
Now, plugging in (\ref{delGs_start}--\ref{delGs_end}) and the corresponding
normalizations, for each line of Eqs.~(\ref{gamma1}) and (\ref{gamma2}) we
get an expression of the following structure: 
\begin{equation} \label{line}
  {\cal N}^{(\kappa_1,\kappa_2,\kappa_3)}\,
  \left[
    \mbox{Tr}_L(\mbox{power series in}\ {\cal F})
    +
    \mbox{Tr}_L(\mbox{power series in}\ {\cal F})
    \cdot\mbox{Tr}_L(\mbox{power series in}\ {\cal F})
  \right]\ .
\end{equation}
Because of the antisymmetry of ${\cal F}$, only even
powers of ${\cal F}$ contribute to the traces. Furthermore we have
the property
\begin{equation}
  {\cal N}^{(\kappa_1,\kappa_2,\kappa_3)}({\cal F})
  =
  {\cal N}^{(\kappa_1,\kappa_2,\kappa_3)}(-{\cal F})\ ,
\end{equation}
and thus (\ref{line}) is invariant with respect to 
${\cal F}\rightarrow {-\cal F}$. Therefore the
${\cal F}\rightarrow {-\cal F}$ terms in Eqs.~(\ref{gamma1}) and 
(\ref{gamma2}) just give a factor of two.

Using this fact, the expressions (\ref{delGs_start}--\ref{delGs_end}) and 
the corresponding normalizations we obtain from 
Eqs.~\eq{gamma1} and \eq{gamma2} straightforwardly:
\beqa{G2sin}
\lefteqn{ \hspace{-10mm}
\Gamma_2[{\cal A}]  = -\frac{1}{2} \,(4\pi)^{-D}
\int\limits_0^{\infty} \!\! dT_1 dT_2 dT_3\; \mbox{det}^{1/2}_L 
\left(\frac{{\cal F}^2}{\Delta^{(2)}_{\cal F}}\right) \Biggl\{ } \nn\\
& & \mbox{Tr}_L \biggl(
  \frac{{\cal F}^2 T_2}{\Delta^{(2)}_{\cal F} }
\Bigl[\, -2\cos{\cal F}(T_1-T_3) \cos2{\cal F}(T_1+T_3)
+\cos{\cal F}(T_3-T_1)  \nn\\
&&+\cos{\cal F} (T_1-T_3)
\mbox{Tr}_L\cos2{\cal F}(T_1+T_3)\,\Bigr]\,\biggr)\nn \\
& &+\,2\mbox{Tr}_L \biggl(
 {{\cal F}\sin{\cal F}T_1\over\Delta^{(2)}_{\cal F}} \Bigl[\,  
 2\sin{\cal F} T_3 \sin2{\cal F}(T_1+T_3) 
-\cos{\cal F}(2T_1-T_3) \nn\\
&& +\,\cos{\cal F}(2T_1+T_3)
\mbox{Tr}_L\cos2{\cal F}T_3\, \Bigr] \,\biggr)\,\Biggr\}
\int\!\!d^D\!x_0\ , 
\eeqa
and
\beqa{G1sin}
\lefteqn{
\Gamma_1[{\cal A}]  = -\frac{1}{2} \,(4\pi)^{-D}
\int\limits_0^{\infty} \!\! dT_1 dT_2 dT_3\; \mbox{det}^{1/2}_L 
\left(\frac{{\cal F}^2}{\Delta^{(2)}_{\cal F}}\right) \Biggl\{ } \nn\\
& & \mbox{Tr}_L\biggl(\, {{\cal F}\over\sin{\cal F}T_3}\biggl[
-2\cos{\cal F}(2T_1+3T_3) +\cos{\cal F}(2T_1+T_3) 
\mbox{Tr}_L\cos 2{\cal F}T_3 \nn\\
&& +\,2\cos{\cal F}T_3\mbox{Tr}_L\cos2{\cal F}(T_1+T_3)
\biggr] \,\biggr) \nn\\
& & +\,\mbox{Tr}_L \biggl(  ({\sin{\cal F}T_1\sin{\cal F}T_3
 \over\Delta^{(2)}_{\cal F}T_2}-{1\over T_2}) \Bigl[\, 
\cos2{\cal F}(T_3-T_1) -\cos2{\cal F}T_3
\mbox{Tr}_L\cos2{\cal F}T_1\,\Bigr]\,\biggr)\nn\\
& & +\,\mbox{Tr}_L\biggl( \, {{\cal F}^2T_2\sin{\cal F}T_1 
\over\Delta^{(2)}_{\cal F}\sin{\cal F}T_3}
\Bigl[\, 2\cos2{\cal F}(T_1+2T_3)-\cos2{\cal F}(T_1+T_3)
\mbox{Tr}_L\cos2{\cal F}T_3 \nn\\   & &-\,\cos2{\cal F}T_3
\mbox{Tr}_L\cos2{\cal F}(T_1+T_3)\,\Bigr] \,\biggr) 
+\,\delta (T_3)\,2 (1-D)\,\mbox{Tr}_L 
\cos 2{\cal F}T_1 \nn\\
& &+\,\delta (T_2)\Bigl[\,
\mbox{Tr}_L \cos 2{\cal F}(T_1-T_3)
-\mbox{Tr}_L (\cos 2{\cal F}T_1)\cdot
 \mbox{Tr}_L (\cos 2{\cal F}T_3)\,\Bigr]\, \Biggr\}
\int\!\!d^D\!x_0\ , 
\eeqa
with
\beq{delb}
 \Delta^{(2)}_{\cal F} = \sin{\cal F}T_1\sin{\cal F}T_3
               + {\cal F}T_2\sin{\cal F}(T_1+T_3)\ .
\eeq
The sum of \eq{G2sin} and \eq{G1sin} gives $I_1[{\cal A}]$ by 
definition, and is certainly equivalent to Eq.~\eq{I1int}. 
In order to obtain Eq.~\eq{I1int} itself,
we should further take the following modification into account. 
Exchanging $T_2$ and $T_3$, and using the relations followed 
from \eq{delb}
\beq{:0}
{{\cal F}\over\sin{\cal F}T_3}=
{{\cal F}\over\Delta^{(2)}_{\cal F}}\sin{\cal F}T_1+
{{\cal F}^2T_2\sin{\cal F}(T_1+T_3)\over\Delta^{(2)}_{\cal F}
\sin{\cal F}T_3}\ ,
\eeq
\beq{:1}
{1\over T_2}-
 {\sin{\cal F}T_1\sin{\cal F}T_3\over\Delta^{(2)}_{\cal F}T_2}
={{\cal F}\over\Delta^{(2)}_{\cal F}}\sin{\cal F}(T_1+T_3)\ ,
\eeq
we finally arrive at the full expression shown in Eq.~\eq{I1int}.

\section{Computational details of $C_i$}\label{ap3}
\setcounter{section}{3}
\setcounter{equation}{0}
\indent

In this Appendix, we show the details of how to perform all the 
integrals in $C_1$, $C_2$, $C_3$ and $C_4$. Let us first perform 
the $T_3$ integrals in $C_i$. Applying the following formula 
to the $T_3$ parts in \eq{Ci}
\beq{form1}
\int_0^\infty e^{-pt}(1+at)^{-\nu}dt = 
p^{\nu-1} a^{-\nu} e^{p/a}\,\Gamma(1-\nu; p/a)\ , 
\qquad  (a>0)
\eeq
and then transforming the $T_1$ and $T_2$ integrals with 
\beq{trans}
\int_0^\infty dT_1 dT_2\, f(T_1,T_2) =
\int_0^\infty dT T\int_0^1 du f(T(1-u),Tu)\ , 
\eeq
we obtain the followings for $i=1$ and 2 (further using $m^2T=t$)
\beq{C12}
C_i = (m^2)^{-2\eps}\int_0^\infty dt \int_0^1 du f_i(u)
t^{2+\eps}e^{-t+tu(1-u)} \Gamma(\eps-3; tu(1-u)), 
\qquad (i=1,2)
\eeq
and for $i=3$ and 4
\beqa{C34}
C_i&=&
-(m^2)^{-2\eps}\int_0^\infty dt \int_0^1 du  f_i(u)t^\eps
e^{-t+tu(1-u)} \Bigl[\,(3-\eps)t \Gamma(\eps-3; tu(1-u)) \nn\\
&&+t^2u(1-u)\Gamma(\eps-3; tu(1-u)) + t^2u(1-u)
{\der\over\der\xi}\Gamma(\eps-3;\xi)\,\Bigr]\ ,
\eeqa
where we have defined
\beq{xi}
            \xi=tu(1-u)
\eeq
and
\beq{f1234}
f_i(u)=u^{i-2a}(1-u)^{a+5-i},\qquad a=\Bigl[{i-1\over2}\Bigr]_G
\eeq
with Gauss' integer symbol $[{\phantom{C}}]_G$. 
In the following, we evaluate \eq{C12} and \eq{C34} separately 
because we shall proceed on different technique and formulae. 

\subsection{$C_1$ and $C_2$}
\indent

We now consider the $t$ integration in \eq{C12}. In the first place, 
it can be integrated in terms of the formula 
(6.455.1 in~\cite{Grad}): 
\beqa{form2}
&&\int_0^\infty t^{\mu-1}e^{-pt}\Gamma(\nu,\alpha t)\,dt =
{\alpha^\nu \Gamma(\mu+\nu)\over\mu(\alpha+p)^{\mu+\nu}}
{}~F(1,\mu+\nu;\mu+1;{p\over\alpha+p}), \nn\\
&&\mbox{Re}\,(\alpha+p), \mbox{Re}\,\mu,\mbox{Re}\,(\mu+\nu)>0\ .
\eeqa
Then applying the formula (9.131.2 in \cite{Grad}),
\beqa{form3}
{}~F(\alpha,\beta;\gamma;z) &=&
{\Gamma(\alpha+\beta-\gamma)\Gamma(\gamma)\over
\Gamma(\alpha)\Gamma(\beta)}(1-z)^{\gamma-\alpha-\beta}
{}~F(\gamma-\alpha,\gamma-\beta;\gamma-\alpha-\beta+1;1-z) \nn\\
&+& {\Gamma(\gamma)\Gamma(\gamma-\alpha-\beta)\over
\Gamma(\gamma-\alpha)\Gamma(\gamma-\beta)}
{}~F(\alpha,\beta;\alpha+\beta-\gamma+1;1-z),
\eeqa
we have 
\beqa{C12y}
C_i &=& (m^2)^{-2\eps} \int_0^1 du f_i(u) 
\Gamma(\eps+3) \Gamma(\eps-3) F(4-\eps,3+\eps;4-\eps;u(1-u)) \nn\\
&+& (m^2)^{-2\eps}\int_0^1 du f_i(u) \Bigl[u(1-u)\Bigr]^{\eps-3} 
{\Gamma(2\eps)\over 3-\eps} F(2\eps,1;\eps-2;u(1-u))\ .
\eeqa
In order to perform the $u$ integrations, we consider the following 
change of variable: cutting the integration region $(0,1)$ in half, 
define $u_+$ ($u_-$) for larger (smaller) values of $u$, and  
apply 
\beq{upm} 
u_\pm = {1\over2}(1\pm\sqrt{1-y}) \ .
\eeq
With this change of variable, the following formula holds 
for an arbitrary function $H(u)$
\beq{utrans}
\int_0^1 du f_i(u) H(u(1-u)) = {1\over4}\int_0^1{dy\over\sqrt{1-y}}
g_i(y) H(y/4)\ ,
\eeq
where we have defined $g_i(y)$ for all $i$:
\beq{gi}
g_i(y)\equiv  f_i\Bigl({1+\sqrt{1-y}\over2}\Bigr)
+f_i\Bigl({1-\sqrt{1-y}\over2}\Bigr)\ , 
\eeq
and we hence have 
\beqa{g12,34}
&& g_1(y)={1\over16}y(4-3y),\qquad g_2(y)={1\over16}y^2, \\
&& g_3(y)={1\over8}y(2-y), \qquad g_4(y)={1\over8}y^2.
\label{g34}
\eeqa
At a glance, one may realize that the $y$ integration can be 
performed by the formula (7.512.12 \cite{Grad})
\beqa{form4}
&&\int_0^1 (1-t)^{\mu-1} t^{\nu-1} F(\alpha,\beta;\gamma;zt)\,dt= 
\mbox{B}(\mu,\nu)\, {}_3F_2(\alpha,\beta,\nu;\gamma,\mu+\nu;z),\nn\\
&&(\mbox{Re}\,\mu, \mbox{Re}\,\nu>0, |z|<1)
\eeqa
and this is exactly the way we obtain the generalized hypergeometric 
${}_3F_2$ expressions \eq{C1F32} and \eq{C2F32}. 

In order to further derive hypergeometric ${}_2F_1$ expressions, 
we rather notice the special case $\gamma=\nu$ in~\eq{form4}. 
Before applying \eq{form4}, we adjust auxiliary variables of 
${}_2F_1$ until the special case is applicable, with using the 
formula (9.137.18 \cite{Grad})
\beq{form5}
{}~F(\alpha,\beta;\gamma;z)=
{\gamma-\alpha\over\gamma}F(\alpha,\beta;\gamma+1;z)
+{\alpha\over\gamma}F(\alpha+1,\beta;\gamma+1;z)\ .
\eeq
Before writing down the final results for $C_1$ and $C_2$, we here 
remark on the way how $C_1$ contains the $\eps'$ dependence. Using 
the following formula (9.137.11 ~\cite{Grad}) with $\beta=0$ 
\beq{form6}
\gamma F(\alpha,\beta;\gamma;z)-
\gamma F(\alpha,\beta+1;\gamma;z)+
\alpha z F(\alpha+1,\beta+1;\gamma+1;z)=0\ ,
\eeq
and then decomposing the second term in \eq{C12y}
\beq{:2}
{}~F(2\eps,1;\eps-2;{y\over4})=1+{2\eps\over\eps-2}\,{y\over4}
{}~F(2\eps+1,1;\eps-1;{y\over4}) \ ,
\eeq
we extract the following integral from $C_1$:
\beqa{l1}
l_1 &\define& 4^{-\eps}\Gamma(2\eps)\int_0^1y^{\eps-2}
(4-3y)(1-y)^{-1/2} \,dy \nn\\
&=& \Gamma(2\eps)\int_0^1 u^{\eps-2}(1-u)^{\eps+1}\,du 
\,=\, \Gamma(2\eps)\mbox{B}(\eps'-1,\eps+2)\ .
\eeqa
Evaluating the $l_1$ contribution separately, we obtain the 
results written in terms of ${}_2F_1$ only: 
\beqa{CiF21}
C_1 &=& (m^2)^{-2\eps} {\Gamma(2\eps)\over3-\eps}
\mbox{B}(\eps'-1,\eps+2) + (m^2)^{-2\eps}{1\over64}
\Gamma(\eps+3)\Gamma(\eps-3) \nn\\
& &\times\Bigl[\, 4\mbox{B}(2,{1\over2})
  {}~F(2,3+\eps;{5\over2};{1\over4}) - 3\mbox{B}(3,{1\over2})
  {}~F(3,3+\eps;{7\over2};{1\over4})\,\Bigr] \nn\\
&+& (m^2)^{-2\eps}4^{-e}{\Gamma(2\eps+1)
   \mbox{B}(\eps,{1\over2})\over(3-\eps)(\eps-2)(\eps-1)}\nn\\
& &\times\Bigl[\,(\eps-2)F(1,2\eps+1;\eps+{1\over2};{1\over4})
  +F(2,2\eps+1;\eps+{1\over2};{1\over4})\,\Bigr] \nn\\
&-& (m^2)^{-2\eps}4^{-e}{3\over4} {\Gamma(2\eps+1)
   \mbox{B}(\eps+1,{1\over2})\over(3-\eps)(\eps-2)}\Bigl[\,
{\eps-2\over\eps}F(1,2\eps+1;\eps+{3\over2};{1\over4}) \nn\\ 
& &+{2(\eps-2)\over\eps(\eps-1)}
{}~F(2,2\eps+1;\eps+{3\over2};{1\over4}) 
  +{2\over\eps(\eps-1)}F(3,2\eps+1;\eps+{3\over2};{1\over4}) 
   \,\Bigr],   \label{C1F21}\\
C_2 &=&(m^2)^{-2\eps}{1\over64}\Gamma(3+\eps)\Gamma(\eps-3) 
\mbox{B}(3,{1\over2}) F(3,3+\eps;{7\over2};{1\over4}) \nn\\
&+&(m^2)^{-2\eps}4^{-\eps}{\Gamma(2\eps)\mbox{B}(\eps,{1\over2})
   \over 1-\eps} \Bigl[\, 
{}~F(1,2\eps;\eps+{1\over2};{1\over4}) \nn\\
& &-{2\over2-\eps} F(2,2\eps;\eps+{1\over2};{1\over4}) 
   +{2\over(2-\eps)(3-\eps)} 
{}~F(3,2\eps;\eps+{1\over2};{1\over4})\,\Bigr]\ . \label{C2F21}
\eeqa

\subsection{$C_3$ and $C_4$}
\indent

In this case, we split $C_i$; $i=3,4$ as 
\beq{R123}
C_i=-(m^2)^{-2\eps}\Bigl[\, (3-\eps)R_1+R_2+R_3\,\Bigr],
\eeq
where 
\beqa{R1,2,3}
R_1&=&\int_0^1du\,f_i(u)\int_0^\infty t^{1+\eps}
e^{-t+tu(1-u)}\Gamma(\eps-3;tu(1-u))\,dt\ , \label{R1} \\
R_2&=&\int_0^1du\,f_i(u)u(1-u)\int_0^\infty t^{2+\eps}
e^{-t+tu(1-u)}\Gamma(\eps-3;tu(1-u))\,dt\ , \label{R2} \\
R_3&=&\int_0^1du\,f_i(u)u(1-u)\int_0^\infty t^{2+\eps}
e^{-t+tu(1-u)}{\der\over\der\xi}\Gamma(\eps-3;\xi)\,dt\ .
\label{R3}
\eeqa
{}~First, we integrate $R_2$ using \eq{form2} 
in the same way as done in \eq{C12}: 
\beq{r2}
R_2={\Gamma(2\eps)\over\eps+3}\int_0^1 f_i(u)
\Bigl[\,u(1-u)\,\Bigr]^{\eps-2} F(1,2\eps;\eps+4;1-u(1-u))\,du.
\eeq
{}~Second, we notice that the incomplete gamma function is 
related to the Whittaker function: 
\beq{form7}
\Gamma(\nu,z) = z^{\nu-1\over2} e^{-z/2} 
W_{{\nu-1\over2},{\nu\over2}}(z),
\eeq
and $W_{\kappa,\nu}(z)$ satisfy
\beqa{form8,9,10}
&&W_{\kappa,\mu}(z)=W_{\kappa,-\mu}(z), \label{form8}\\
&&z\der_z W_{\kappa,\mu}(z)=
\Bigl({z\over2}-\kappa \Bigr)W_{\kappa,\mu}(z)
-W_{\kappa+1,\mu}(z), \label{form9}\\
&&W_{\kappa,\mu}(z)=z^{1/2}W_{\kappa-1/2,\mu+1/2}(z) 
+\Bigl({1\over2}-\kappa-\mu\Bigr)W_{\kappa-1,\mu}(z)\ .
\label{form10}
\eeqa
Applying the formula \eq{form7} and its derivative to $R_1$ 
and $R_3$, one can prove the relation
\beq{R3toR4}
R_3 ={\eps-4\over2}R_1 -{1\over2}R_2 + R_4 \ ,
\eeq
where 
\beq{R4}
R_4 =\int_0^1 du f_i(u)\Bigl[u(1-u)\Bigr]^{\eps-2\over2}
\int_0^\infty t^{3\eps/2} e^{-t+tu(1-u)/2}
\der_\xi W_{{\eps-4\over2},{\eps-3\over2}}(\xi)\,dt\ .
\eeq
We here remove the derivative $\der_\xi W$ from the r.h.s. of 
Eq.\eq{R4}, making use of the relation~\eq{form9}: 
\beq{R4b}
R_4 = {1\over2}R_2 - {\eps-4\over2}R_1 -
- \int_0^1 du f_i(u) \Bigl[u(1-u)\Bigr]^{\eps-4\over2}
\int_0^\infty  t^{{3\over2}\eps-1} e^{-t+tu(1-u)/2}
W_{{\eps-2\over2},{\eps-3\over2}}(\xi)\,dt\ .
\eeq 
Adjusting the indices on $W$ on the r.h.s. of \eq{R4b} in terms 
of the recursion relation~\eq{form10} (with $\kappa=(\eps-2)/2,
\mu=(\eps-3)/2$), we then apply the formula
\beqa{form11}
\int_0^\infty t^\nu W_{\kappa,\mu}(at)e^{-pt}\,dt &=&
{\Gamma(\nu+\mu+{3\over2})\Gamma(\nu-\mu+{3\over2})a^{\mu+1/2}
\over\Gamma(\nu-\kappa+2)(p+a/2)^{\nu+\mu+3/2}} \nn\\
&&\times F\Bigl(\nu+\mu+{3\over2},\mu-\kappa+{1\over2};
\nu-\kappa+2;{2p-a\over2p+a}\Bigr), \nn\\
&&(\mbox{Re}\,\nu+{3\over2}-|\mbox{Re}\,\mu|>0)
\eeqa
and the r.h.s. of \eq{R4b} becomes
\beq{R4bb} 
 {1\over2}R_2 - {\eps-4\over2}R_1 -(3-\eps)R_1
-{\Gamma(2\eps)\Gamma(\eps+2)\over\Gamma(\eps+3)}
\int_0^1 f_i(u)\Bigl[u(1-u)\Bigr]^{\eps-2}
{}~F(2\eps,1;\eps+3;1-u(1-u))\ .
\eeq
Substituting this expression for $R_4$ in Eq.\eq{R3toR4}, which should 
then be inserted into Eq.\eq{R123}, and thereby eliminating $R_1$, we 
calculate $C_i$ as follows:
\beqa{C34u} 
C_i&=&(m^2)^{-2\eps}\int_0^1 du\,
f_i(u)\Bigl[u(1-u)\Bigr]^{\eps-2}
\Gamma(2\eps)\Bigl[{-1\over3+\eps}F(1,2\eps;\eps+4;1-u(1-u))\nn\\
&+&{1\over2+\eps}F(1,2\eps;\eps+3;1-u(1-u))\,\Bigr] \nn\\
&=& {(m^2)^{-2\eps}\Gamma(2\eps)\over4(2+\eps)(3+\eps)}
\int_0^1 g_i(y)(1-y)^{-{1\over2}}({y\over4})^{\eps-2}
{}~F(2,2\eps;\eps+4;1-{y\over4})\, dy\ ,
\eeqa
where we have used the transformations \eq{utrans} and \eq{form5} 
at the second equality. We now transform the argument of the 
hypergeometric function from $1-y/4$ to $y/4$ through the 
formula~\eq{form3}, and thus giving rise to 
$F(4-\eps,2+\eps;3-\eps;y/4)$ and $F(2,2\eps;\eps-1;y/4)$. 
(If we apply \eq{form4} at this stage, we obtain the generalized 
hypergeometric ${}_3F_2$ expressions \eq{C3F32} and \eq{C4F32}.) 

{}~For the purpose to reduce ${}_3F_2$ to ${}_2F_1$, we 
apply \eq{form5} twice to the latter $F$ (two of them explained 
right above), and apply the following formula to the former $F$ 
(setting $\alpha=\gamma=3-\eps,\beta=2+\eps$):
\beq{form12}
{}~F(\alpha+1,\beta;\gamma;z)=
{\beta\over\alpha}F(\alpha,\beta+1;\gamma;z)
+{\alpha-\beta\over\alpha}F(\alpha,\beta;\gamma;z)\ .
\eeq
After that, we are able to integrate them by using \eq{form4} 
(with $\gamma=\nu$), and the results are therefore 
\beqa{CjF21}
C_3&=&(m^2)^{-2\eps}{1\over16}{\Gamma(4+\eps)\Gamma(\eps-2)
\over(\eps+2)(\eps+3)(3-\eps)} \mbox{B}(2,{1\over2}) 
\Bigl[\, (2+\eps)F(2,3+\eps;{5\over2};{1\over4}) \nn\\
&&+(1-2\eps)F(2,2+\eps;{5\over2};{1\over4})\,\Bigr]\nn\\
&+&(m^2)^{-2\eps}\Gamma(2\eps){\Gamma(2-\eps)\over\Gamma(4-\eps)}
{\mbox{B}({1\over2},\eps)\over(\eps-1)4^\eps}
\Bigl[\, (\eps-3)F(2,2\eps;\eps+{1\over2};{1\over4})\nn\\
&&+2F(3,2\eps;\eps+{1\over2};{1\over4})\,\Bigr] -C_4\ ,
\label{C3F21}\\
C_4&=&(m^2)^{-2\eps}{1\over32}{\Gamma(4+\eps)\Gamma(\eps-2)
\over(\eps+2)(\eps+3)(3-\eps)} \mbox{B}(3,{1\over2}) 
\Bigl[\, (2+\eps)F(3,3+\eps;{7\over2};{1\over4}) \nn\\
&&+(1-2\eps)F(3,2+\eps;{7\over2};{1\over4})\,\Bigr]\nn\\
&+&(m^2)^{-2\eps}\Gamma(2\eps){\Gamma(2-\eps)\over\Gamma(4-\eps)}
{\mbox{B}({1\over2},\eps+1)\over2\eps(\eps-1)4^\eps}
\Bigl[\, (\eps-3)(\eps-2)F(2,2\eps;\eps+{3\over2};{1\over4})\nn\\
&&+4(\eps-3)F(3,2\eps;\eps+{3\over2};{1\over4})
+6F(4,2\eps;\eps+{3\over2};{1\over4})\,\Bigr]\ . \label{C4F21}
\eeqa

\section{The proof of Eq.(5.14)}\label{ap4}
\setcounter{section}{4}
\setcounter{equation}{0}
\indent

We show the outline of how to prove the formula \eq{Form4}. Basically 
we follow the same technique as we used in Appendix~\ref{ap3}, 
with using \eq{form4} and \eq{form5} for the aim of rearranging 
${}_3F_2$ into ${}_2F_1$. We evaluate the l.h.s. of Eq.~\eq{form4} in 
two ways: one is the direct result (the r.h.s. of the formula), and 
the other is a combination with \eq{form5}. 

On the one hand, setting $\gamma=\nu-n$ in \eq{form4}, we have
\beqa{lhs}
&&\int_0^1 (1-t)^{\mu-1} t^{\nu-1} F(\alpha,\beta;\nu-n;zt)\,dt= 
\mbox{B}(\mu,\nu)\, {}_3F_2(\alpha,\beta,\nu;\nu-n,\mu+\nu;z),\nn\\
&&(\mbox{Re}\,\mu, \mbox{Re}\,\nu>0, |z|<1) \ .
\eeqa
On the other hand, using \eq{form5} $n\geq 1$  times, the integrand 
in Eq. \eq{lhs} can be expanded as 
\beq{inte}
(1-t)^{\mu-1}t^{\nu-1}
\prod_{k=1}^n{\nu-\alpha-k\over\nu-k} \sum_{r=0}^n
\prod_{p=1}^r{\alpha+p-1\over\nu-\alpha-p}{n\choose r}
{}~F(\alpha+r,\beta;\nu;zt)\ ,
\eeq 
where we define $\prod_{p=1}^0\equiv1$. Then applying~\eq{form4} 
(with the case $\gamma=\nu$) to this expression, 
the l.h.s. of \eq{lhs} becomes
\beq{lhs2}
\mbox{B}(\mu,\nu)
\prod_{k=1}^n{\nu-\alpha-k\over\nu-k} \sum_{r=0}^n
\prod_{p=1}^r{\alpha+p-1\over\nu-\alpha-p}{n\choose r}
{}~F(\alpha+r,\beta;\mu+\nu;z)\ .
\eeq 
We thus have the equality
\beq{ltor}
{}_3F_2(\alpha,\beta,\nu;\nu-n,\mu+\nu;z)=
\prod_{k=1}^n {\nu-\alpha-k\over\nu-k} \sum_{r=0}^n
\prod_{p=1}^r{\alpha+p-1\over\nu-\alpha-p}{n\choose r}
{}~F(\alpha+r,\beta;\mu+\nu;z)\ .
\eeq
Putting $\alpha=1$, $\nu=n+2$, and using the identity
\beq{note}
\prod_{p=1}^r {p\over n+1-p} {n\choose r}=1\ ,
\eeq
we therefore have proven the formula \eq{Form4}:
\beq{proven}
{}_3F_2(1,\beta,n+2;2,\mu+\nu;z)=
{1\over n+1} \sum_{r=0}^n
{}~F(r+1,\beta;\mu+\nu;z)\ ,
\eeq
which holds for $\Re(\mu), \Re(\nu) > 0$, $|z|<1$ and $n\geq0$ 
(we have proven it for $n\geq 1$, but the $n=0$ case is trivial).

\section{Feynman diagram results}\label{ap5}
\setcounter{section}{5}
\setcounter{equation}{0}
\indent

In this Appendix, we present some Feynman diagram calculations 
in reference to the results obtained by our method. We follow the 
same notations as Refs.~\cite{Abo,CM} in the Minkowski space, 
however use the massive propagator in the (background) Feynman gauge:

\vspace{3mm}
\begin{minipage}[h]{15cm} 
\begin{center}
\input{wave.pictex}
\end{center}
\end{minipage}
\vspace{3mm}

We only deal with the parts which contain the ``eight-figure'' 
vacuum diagram. When the massive propagator is introduced, the 
tadpole contributions remain (see the diagrams (a) and (b) in 
{}~Figure 2). After some calculations, the tadpole diagram (a) reads  
\beq{Tada}
{\Pi_T^{(a)}}^{ab}_{\mu\nu}=g^4{C_A^2\delta^{ab}\over(4\pi)^{2-\eps}}
(m^2)^{1-\eps}(3-2\eps)\Gamma(\eps-1)\Bigl[\,
8(k^2g_{\mu\nu}-k_\mu k_\nu)J_1 + 
D(k_\mu k_\nu J_1  + 4k_\nu J_2  +  4 J_3)\,\Bigr]\ ,
\eeq
where
\beqa{Ji}
&&J_1=\int {d^Dp\over(2\pi)^Di}{1\over((p+k)^2-m^2)(p^2-m^2)^2},\\
&&J_2=\int {d^Dp\over(2\pi)^Di}{p_\mu\over((p+k)^2-m^2)(p^2-m^2)^2},\\
&&J_3=\int {d^Dp\over(2\pi)^Di}{p_\mu p_\nu\over((p+k)^2-m^2)(p^2-m^2)^2},
\eeqa
and these are equal to 
\beqa{Jiji}
&&J_1=-{\Gamma(\eps+1)\over(4\pi)^{2-\eps}}\, j_1\ ,\qquad
J_2={\Gamma(\eps+1)\over(4\pi)^{2-\eps}} k_\mu\, j_2\ ,\nn\\
&&J_3={\Gamma(\eps+1)\over(4\pi)^{2-\eps}}
({1\over2\eps}g_{\mu\nu}\,j_3-{1\over2}k_\mu k_\nu\,j_2)\ ,
\eeqa
with
\beqa{j1,2,3}
&&j_1= \int_0^1 x\Bigl(m^2-k^2(x-x^2)\Bigr)^{-1-\eps}\,dx, \\
&&j_2= \int_0^1 x(1-x)\Bigl(m^2-k^2(x-x^2)\Bigr)^{-1-\eps}\,dx,\\
&&j_3= \int_0^1 x\Bigl(m^2-k^2(x-x^2)\Bigr)^{-\eps}\, =\, 
{1\over2}(m^2)^{-\eps}-{1\over2}\eps k^2(2j_2-j_1)\ .\label{j3}
\eeqa
Thus Eq.\eq{Tada} becomes
\beqa{Tada2}
{\Pi_T^{(a)}}^{ab}_{\mu\nu}&=&g^4{C_A^2\delta^{ab}\over(4\pi)^{4-2\eps}}
(m^2)^{1-\eps}(3-2\eps)\Gamma(\eps-1)\Gamma(\eps+1)\Bigl[\,
{D\over\eps}(m^2)^{-\eps} \nn\\ &&+
(k^2g_{\mu\nu}-k_\mu k_\nu)\Bigl(\, (D-8)j_1-2Dj_2,\Bigr) 
\,\Bigr]\ ,
\eeqa
and the other tadpole diagram (b) is calculated as
\beq{Tadb}
{\Pi_T^{(b)}}^{ab}_{\mu\nu}=g^4{C_A^2\delta^{ab}\over(4\pi)^{4-2\eps}}
(m^2)^{1-2\eps}g_{\mu\nu}(3-2\eps)D\Gamma(\eps-1)\Gamma(\eps)\ .
\eeq
As a result, the sum of these tadpole contributions takes the 
transversal form:
\beqa{Tsum}
{\Pi_T}^{ab}_{\mu\nu}&=&
{\Pi_T^{(a)}}^{ab}_{\mu\nu}+{\Pi_T^{(b)}}^{ab}_{\mu\nu}\nn\\
&=& g^4{C_A^2\delta^{ab}\over(4\pi)^{4-2\eps}} 
(m^2)^{1-\eps}(3-2\eps)\Gamma(\eps-1)\Gamma(\eps+1)\nn\\
&&\times
(k^2g_{\mu\nu}-k_\mu k_\nu)\Bigl[\, (D-8)j_1-2Dj_2\,\Bigr]\ . 
\eeqa
When $k^2\ra0$ ($|k^2|<<m^2$), the quantities $j_i$ behave as 
\beq{jik0}
j_1={1\over2}(m^2)^{-1-\eps}\ ,\qquad
j_2={1\over6}(m^2)^{-1-\eps}\ ,\qquad
j_3={1\over2}(m^2)^{-\eps}\ ,
\eeq
and when $m^2\ra0$, they reduce to 
\beqa{jim0}
&&j_1={1\over2}(-k^2)^{-1-\eps}\mbox{B}(-\eps,-\eps)\ ,\qquad
j_2=(-k^2)^{-1-\eps}\mbox{B}(1-\eps,1-\eps)\ ,\nn\\
&&j_3={1\over2}(-k^2)^{-\eps}\mbox{B}(1-\eps,1-\eps)\ .
\eeqa
Therefore in these limits Eq.\eq{Tsum} behaves 
\beq{limPT}
{\Pi_T}^{ab}_{\mu\nu}= g_0^4{C_A^2\delta^{ab}\over(4\pi)^4} 
(k^2g_{\mu\nu}-k_\mu k_\nu)
\left\{ \begin{array}{ll}
{\displaystyle{10\over\eps}}    &\quad (k^2\ra0) \\ 
 \strut     0          &\quad (m=0)   
\end{array}\right.\quad +\,{\cal O}(1)\ .
\eeq

Next, the diagram (c) amounts to 
\beq{Pie}
{\Pi^{(e)}}^{ab}_{\mu\nu}=-6
g_0^4{C_A^2\delta^{ab}\over(4\pi)^4} 
(k^2g_{\mu\nu}-k_\mu k_\nu)\exp[2\eps\ln(4\pi\mu^2)]
\Gamma^2(\eps)(2j_3)^2\ ,
\eeq
where $j_3$ is given by \eq{j3}. Substituting Eqs.~\eq{jik0} and 
\eq{jim0} for this $j_3$ in each limit, we derive the following 
limits of \eq{Pie}:
\beq{limPe}
{\Pi^{(e)}}^{ab}_{\mu\nu}= g_0^4{C_A^2\delta^{ab}\over(4\pi)^4} 
(k^2g_{\mu\nu}-k_\mu k_\nu)
\left\{ \begin{array}{ll} 
{\displaystyle
-{6\over\eps^2}+{12\over\eps}\rho_m}    &\quad (k^2\ra0) \\
\lower10pt\hbox{${\displaystyle       
-{6\over\eps^2}-{24\over\eps}+{12\over\eps}\rho}$}  &\quad 
\lower10pt\hbox{$(m=0)$}   
\end{array}\right.\quad +\,{\cal O}(1)\ ,
\eeq
where $\rho_m$ is given by \eq{rhom}, and 
\beq{rho}
\rho=\gamma_{\scriptscriptstyle{E}}+\ln{-k^2\over4\pi\mu^2}\ .
\eeq
We therefore conclude that our results coincide with 
the Feynman diagram calculations in the situation $k^2\ra0$ 
($m\not= 0$) (q.v. \eq{pi5} and \eq{pi6}):
\beqa{pipi}
&&{\Pi_T}^{ab}_{\mu\nu}=
g_0^4{C_A^2\delta^{ab}\over(4\pi)^4} 
(k^2g_{\mu\nu}-k_\mu k_\nu)(-10 C'_6) = {\Pi_6}^{ab}_{\mu\nu}\ , \\
&&{\Pi^{(e)}}^{ab}_{\mu\nu}=
g_0^4{C_A^2\delta^{ab}\over(4\pi)^4} 
(k^2g_{\mu\nu}-k_\mu k_\nu)(-6 C'_5) = {\Pi_5}^{ab}_{\mu\nu}\ .
\eeqa

\end{appendix}


\end{document}